%% file: IDM_arxiv_v2.tex
\documentclass[a4paper, 12pt, oneside]{article}
\pdfoutput=1
\usepackage{jheppub}
\makeatletter
\def\@fpheader{}
\makeatother

\usepackage{graphicx,wrapfig,float,slashed,subcaption}
\usepackage{amsmath,amssymb,epsfig,graphicx,xcolor}
\usepackage{epstopdf}
\usepackage{multirow}

\usepackage{ragged2e}
\usepackage{lipsum}
\usepackage{array}

\usepackage{tikz}
\usetikzlibrary{arrows,positioning,shapes}
\usetikzlibrary{decorations.pathmorphing}
\usetikzlibrary{decorations.markings}

\usepackage{cleveref}

\newcommand{\nc}{\newcommand}
\nc{\non}{\nonumber}
\nc{\hc}{\hbox {H.c.}}
\nc{\noi}{\noinde	nt}
\nc{\barx}{\bar{x}}
\nc{\pbarn}{\;\hbox {pb}}
\nc{\fbarn}{\;\hbox {fb}}
\nc{\ra}{\rightarrow}
\nc{\met}{p_{T}^{\textnormal{miss}}}
\nc{\lep}{\ell}
\usepackage{color}

\definecolor{agray}{rgb}{0.95, 0.95, 0.99}
\nc{\hsp}{\hspace{0.5cm}}
\nc{\lsp}{\hspace{1cm}}
\nc{\Lsp}{\hspace{2cm}}
\nc{\LLsp}{\lsp\lsp}
\nc{\lra}{\longrightarrow}
\nc{\p}{\prime}
\nc{\sgn}{\text{sgn}}
\nc{\ph}{\varphi}
\nc{\beq}{\begin{equation}}  \nc{\eeq}{\end{equation}}
\nc{\bea}{\begin{eqnarray}}  \nc{\eea}{\end{eqnarray}}
\nc{\baa}{\begin{array}}     \nc{\eaa}{\end{array}}
\nc{\bit}{\begin{itemize}}   \nc{\eit}{\end{itemize}}
\nc{\ben}{\begin{enumerate}} \nc{\een}{\end{enumerate}}
\nc{\bce}{\begin{center}}    \nc{\ece}{\end{center}}
\nc{\bpm}{\begin{pmatrix}}   \nc{\epm}{\end{pmatrix}}
\nc{\bvt}{\begin{verbatim}}  \nc{\evt}{\end{verbatim}}

\def\lsim{\mathrel{\raise.3ex\hbox{$<$\kern-.75em\lower1ex\hbox{$\sim$}}}}
\def\gsim{\mathrel{\raise.3ex\hbox{$>$\kern-.75em\lower1ex\hbox{$\sim$}}}}

\def\udots{\mathinner{\mkern1mu\raise1pt\vbox{\kern7pt\hbox{.}}\mkern2mu\raise4pt\hbox{.}\mkern2mu\raise7pt\hbox{.}\mkern1mu}}

\newcommand\fverb{\setbox\fverbbox=\hbox\bgroup\verb}
\newcommand\fverbdo{\egroup\medskip\noindent%
			\fbox{\unhbox\fverbbox}\ }
\newcommand\fverbit{\egroup\item[\fbox{\unhbox\fverbbox}]}
\newbox\fverbbox

\makeatletter
\renewcommand{\boxed}[1]{\textcolor{black}{%
\tikz[baseline={([yshift=-0ex]current bounding box.center)}] \node [rectangle, minimum width=0ex,draw] {\normalcolor\m@th$\displaystyle#1$};}}
 \makeatother

\title{Production of Inert Scalars at the high energy $e^+ e^-$ colliders}

\author[a]{Majid Hashemi,} 
\author[b]{Maria Krawczyk,}
\author[b]{Saereh Najjari,}
\author[b]{Aleksander~Filip~\.Zarnecki }
\affiliation[a]{Physics Department, College of Sciences,\\
Shiraz University, Shiraz, 71946-84795, Iran}
\affiliation[b]{Faculty of Physics,
University of Warsaw,\\
Pasteura 5, 02-093 Warsaw, Poland}
\emailAdd{hashemi\_mj@shirazu.ac.ir}
\emailAdd{maria.krawczyk@fuw.edu.pl}
\emailAdd{saereh.najjari@fuw.edu.pl}
\emailAdd{zarnecki@fuw.edu.pl}

\abstract{
  We investigate the phenomenology of the light charged and
  neutral scalars in Inert Doublet Model at future $e^+ e^-$
  colliders with center of mass energies of 0.5 and 1 TeV, and
  integrated luminosity of 500~fb$^{-1}$.
  The analysis covers two production processes,
  $e^{+}e^{-} \rightarrow H^{+}H^{-}$ and $e^{+}e^{-}\rightarrow AH$,
  and consists of signal selections, cross section determinations
  as well as dark matter mass measurements.
  Several benchmark points are studied with focus on
  $H^{\pm} \rightarrow W^{\pm}H$ and $A \rightarrow ZH$ decays.
  It is concluded that the signal will be well observable in
  different final states allowing for mass determination of all new scalars with statistical precision of the order of few hundred MeV.
 }

\keywords{Beyond Standard Model, Inert Doublet Model, Inert Scalars,
  \mbox{Dark Matter}, $e^+e^-$ Colliders}

\begin{document}

\maketitle
\flushbottom

\tikzstyle{every picture}+=[remember picture]
\pgfdeclaredecoration{complete sines}{initial}
{
    \state{initial}[
        width=+0pt,
        next state=sine,
        persistent precomputation={\pgfmathsetmacro\matchinglength{
            \pgfdecoratedinputsegmentlength / int(\pgfdecoratedinputsegmentlength/\pgfdecorationsegmentlength)}
            \setlength{\pgfdecorationsegmentlength}{\matchinglength pt}
        }] {}
    \state{sine}[width=\pgfdecorationsegmentlength]{
        \pgfpathsine{\pgfpoint{0.25\pgfdecorationsegmentlength}{0.5\pgfdecorationsegmentamplitude}}
        \pgfpathcosine{\pgfpoint{0.25\pgfdecorationsegmentlength}{-0.5\pgfdecorationsegmentamplitude}}
        \pgfpathsine{\pgfpoint{0.25\pgfdecorationsegmentlength}{-0.5\pgfdecorationsegmentamplitude}}
        \pgfpathcosine{\pgfpoint{0.25\pgfdecorationsegmentlength}{0.5\pgfdecorationsegmentamplitude}}
}
    \state{final}{}
}

\tikzset{
fermion/.style={very thick,draw=black, line cap=round, postaction={decorate},
    decoration={markings,mark=at position .65 with {\arrow[black]{latex}}}},
photon/.style={very thick, line cap=round,decorate, draw=black,
    decoration={complete sines,amplitude=4pt, segment length=8pt}},
boson/.style={very thick, line cap=round,decorate, draw=black,
    decoration={complete sines,amplitude=4pt,segment length=8pt}},
gluon/.style={very thick,line cap=round, decorate, draw=black,
    decoration={coil,aspect=1,amplitude=4pt, segment length=8pt}},
scalar/.style={dashed, very thick,line cap=round, decorate, draw=black},
cscalar/.style={dashed,very  thick,line cap=round, draw=black,postaction={decorate},
    decoration={markings,mark=at position .65 with {\arrow[black]{latex}}}},
jet/.style={double, thick,line cap=round, decorate, draw=black},
ghost/.style={dotted, thick,line cap=round, decorate, draw=black},
->-/.style={decoration={
  markings,
  mark=at position 0.6 with {\arrow{>}}},postaction={decorate}}
 }

 \makeatletter
\tikzset{
    position/.style args={#1 degrees from #2}{
        at=(#2.#1), anchor=#1+180, shift=(#1:\tikz@node@distance)
    }
}
\makeatother



\section{Introduction}
\label{Inroduction}
Inert Doublet Model (IDM) is one of the simplest extensions of
the Standard Model (SM), with an additional $SU(2)$  scalar doublet,
which can provide a dark matter candidate \cite{Deshpande:1977rw,Cao:2007rm,Barbieri:2006dq,LopezHonorez:2006gr,Honorez:2010re,Dolle:2009fn,Goudelis:2013uca,Krawczyk:2013jta}. The
scalar sector of IDM consist of two SU(2) doublets where one is the
SM-like Higgs doublet while the other is the inert or dark doublet. The
scalar sector of the theory respects a discrete $Z_2$ symmetry under
which the SM Higgs doublet $\Phi_S$ is {\it even} (as well as all the
other SM fields) while the inert doublet $\Phi_D$ is {\it odd},
i.e. $\Phi_S\to\Phi_S$  (SM$\to$SM) and $\Phi_D\to-\Phi_D$. Due to the $Z_2$
symmetry, only the SM Higgs doublet acquires a non-zero vacuum
expectation value and hence is a source of electroweak symmetry
breaking (EWSB). After EWSB in the scalar sector this model has five
physical states: the SM Higgs boson $h$ as well as two
charged scalars, $H^\pm$, and two  neutral ones, $H$ and $A$. Since
the inert doublet is odd under $Z_2$ symmetry, the lightest inert particle
is a natural candidate for dark matter. Also due  to the $Z_2$
symmetry, the inert doublet does not couple with the fermions of the
SM through Yukawa-type interactions. { {This model provides
 description of the evolution of the universe \cite{Ginzburg:2010wa} and
 strong first order phase transition, needed for baryogenesis
 \cite{Chowdhury:2011ga,Borah:2012pu,Gil:2012ya,Cline:2013bln}}}.

In this work, we consider scenarios where  $H$ boson is the dark matter
candidate ($m_H <m_{H^\pm}, m_A$), using the benchmark points
suggested in Ref.~\cite{Ilnicka:2015jba}, which satisfy all the
recent experimental and theoretical constraints. We study
the potential of the future $e^+ e^-$  colliders, like ILC or CLIC, for testing
the IDM; other analyses of the IDM at colliders were done
in \cite{Aoki:2013lhm,Ho:2013spa,Lundstrom:2008ai,Dolle:2009ft,Gustafsson:2012aj,Arhrib:2013ela,
Krawczyk:2013jta,Arhrib:2012ia,Belanger:2015kga,Swiezewska:2012eh,
Ginzburg:2014ora}.  We focus on the charged scalar ($H^+\;
H^-$) production and the neutral scalar ($H \;A$) production at the
center of mass energies of 0.5~TeV and 1~TeV, with the integrated
luminosity of 500~fb$^{-1}$ corresponding to the first 4 years of ILC
running~\cite{Barklow:2015tja}.
In particular we consider the following decay processes:
\beq
\begin{split}
  &e^+ e^- \to H^+ H^- \to W^+ W^- HH \to
  \lep \nu j j HH, jjjjHH,
\\
&e^+ e^- \to H A \to H H Z  \to HH \lep \lep, HH jj.
\label{yukawa_lagrangian_2}\end{split}
\eeq
A similar analysis for different benchmark points at the ILC with
center of mass energies of 250~GeV to 500~GeV has been performed in
\cite{Aoki:2013lhm}. { The analysis of the Compressed IDM (with a degenerated spectrum  of inert scalars)
have been recently studied in \cite{Blinov:2015qva} for LEP, as well
as for LHC and ILC.}

The paper is organized as follows. Essential details of our model
setup and the benchmark points are described in
\cref{Inert Doublet Model}. In \cref{Software Setup} we
provide the description of simulation tools used in the
analysis. \Cref{eeHH,eeHA} contain the details of
the event generation and physics analysis of our benchmark points for
$e^+ e^- \to H^+ H^-$ and $e^{+}e^{-}\ra AH$, respectively. {In \cref{sec:dm_mass} we propose a procedure for the measurement of the dark matter mass.} Finally,
the conclusions are given in \cref{Conclusion}.



\section{Inert Doublet Model}
\label{Inert Doublet Model}
The scalar sector of IDM consists of two scalar doublets, the SM Higgs
doublet $\Phi_S$  with SM-like Higgs boson $h$ and the inert doublet
$\Phi_D$. Only the SM Higgs doublet($\Phi_S$) interacts with the SM
fermions, whereas  the inert doublet ($\Phi_D$) is $Z_2$ odd and it
does not interact with the SM fermions through Yukawa-type
interactions. The two doublets can be parameterised as follows,
\beq
\Phi_S=\bpm G^\pm\\ \frac{v+h+iG^0}{\sqrt2} \epm, \hspace{2cm} \Phi_D=\bpm H^\pm\\\frac{H+iA }{\sqrt2} \epm,	\label{higgs_vev}
\eeq
with  the vacuum expectation value $v=246$~GeV (the SM value). The
most general scalar potential for the IDM has the following form:
\begin{align}
V(\Phi_S&,\Phi_D)=-\frac{1}{2}\Big[m_{11}^2(\Phi_S^\dagger\Phi_S)+m_{22}^2(\Phi_D^\dagger\Phi_D)\Big]+\frac{\lambda_1}{2}(\Phi_S^\dagger\Phi_S)^2+\frac{\lambda_2}{2}(\Phi_D^\dagger\Phi_D)^2 \notag\\
&+\lambda_3(\Phi_S^\dagger\Phi_S)(\Phi_D^\dagger\Phi_D)+\lambda_4(\Phi_S^\dagger\Phi_D)(\Phi_D^\dagger\Phi_S)
+\frac{\lambda_5}{2}\Big[(\Phi_S^\dagger\Phi_D)^2+(\Phi_D^\dagger\Phi_S)^2\Big].	\label{potential}
\end{align}
The above potential has seven parameters ($m_{11,22},
\lambda_{1,2,3,4,5}$) that we assume to be real.
The  scalar masses are  as follows:
\begin{align}
m_h^2&=\lambda_1 v^2=m_{11}^2, 	\notag\\
m_{H^+}^2&=\frac{1}{2}(\lambda_3 v^2-m_{22}^2),		\notag\\
m_H^2&=\frac{1}{2}(\lambda_{345}v^2-m_{22}^2), 	\notag\\
m_A^2&=\frac{1}{2}(\bar{\lambda}_{345}v^2-m_{22}^2),	\label{4d_cc}
\end{align}
with $\lambda_{345}\equiv\lambda_3+\lambda_4+\lambda_5$ and
$\bar{\lambda}_{345}\equiv\lambda_3+\lambda_4-\lambda_5$.

\subsection*{Theoretical Constraints}

The scalar potential $V(\Phi_S,\Phi_D)$ \eqref{potential} has to satisfy
many theoretical and experimental constraints, as discussed in
Ref.~\cite{Ilnicka:2015jba}, which we have to take into account
when defining benchmark scenarios.
The vacuum stability at tree level leads to the following conditions on the couplings:
\beq \lambda_1\geq 0,~\lambda_2\geq 0,~\sqrt{\lambda_1\lambda_2}+\lambda_3>0,~~
\sqrt{\lambda_1\lambda_2}+\lambda_{345}>0 \eeq
To have the inert vacuum as a global minimum of the potential,
    we require  \cite{Swiezewska:2012ej}%
\footnote{See also \cite{Swiezewska:2015paa,Ferreira:2015pfi} for more
    detailed discussion}:
  \beq \frac{m_{11}^2}{\sqrt{\lambda_1}}\geq \frac{m_{22}^2}{\sqrt{\lambda_2}}.\eeq
We also require perturbative unitarity of $2 \to 2$ scalar scattering matrix.

\subsection*{Experimental Constraints}

We set the mass of the SM-like Higgs boson $h$ to be $M_h=125$~GeV \cite{ Aad:2015zhl}
and impose the upper bound on the total width of $h$, $\Gamma_{tot}\leq 22$~MeV \cite{Khachatryan:2014iha,Aad:2015xua}.
Total widths of $W$ and $Z$ boson imply the following bounds \cite{Agashe:2014kda}:
\beq m_H+m_A\geq m_Z,~~~~2m_{H^\pm}\geq m_Z ,~~m_A+m_{H_\pm},m_H+m_{H^\pm}\geq m_W. \eeq
We take into account following search results:
direct bounds on the dark matter scattering from LUX experiment~\cite{Akerib:2013tjd},
LEP limit on the charged scalar mass of $m_{H^\pm}\geq 70$~GeV \cite{Pierce:2007ut},
exclusions from SUSY searches at LHC and LEP \cite{Lundstrom:2008ai,Belanger:2015kga}
as well as the limit on the charged scalar width,
$\Gamma_{tot}\geq 6.58 \times 10^{-18}$~GeV \cite{Ilnicka:2015jba}.
Finally, we require the agreement (at 2$\sigma$ level) with electroweak precision
observables~\cite{Altarelli:1990zd,Peskin:1990zt,Maksymyk:1993zm,Peskin:1991sw}
and with upper limit on relic density from Planck measurement, $\Omega_c h^2\leq 0.1241 $ \cite{Ade:2015xua}.

\subsection*{Benchmark Points}
For our collider analysis we consider the set of benchmark
points (BP) proposed in \cite{Ilnicka:2015jba}, which satisfy all
the above mentioned constraints.
These BP scenarios can be difficult for precise measurement at the LHC,
due to small mass differences between new scalars,
but should be clearly visible at the $e^+ e^-$ colliders.  We consider

\def\arraystretch{1.5}
\begin{tabular}{llll}
 BP 1:  \it  &$m_H$= 57.5 GeV,& $m_A$=113 GeV, &$m_{H^\pm}$=123 GeV,\\
 BP 2: \it  &$m_H$= 85.5 GeV,& $m_A$=111 GeV, &$m_{H^\pm}$=140 GeV,\\
 BP 3: \it &$m_H$= 128 GeV,& $m_A$=134 GeV,& $m_{H^\pm}$=176 GeV.
\end{tabular}
\def\arraystretch{1.0}

~

\noindent
Our analysis is limited to the three low mass scenarios. The high mass
scenarios are much more challenging as the production cross
sections for signal events are very low and the observable decay
products have low energies.

\section{Software Setup}
\label{Software Setup}

Signal events, i.e. pair produced charged and neutral scalars in
$e^+e^-$ collisions, are generated using CompHEP 4.5.2
\cite{comphep1,comphep2}. It uses IDM model files which are prepared
using LanHEP 3.2 \cite{lanhep1,lanhep2}. The output of CompHEP in LHEF
(Les Houches Event File) format is passed to PYTHIA 8.1.53
\cite{pythia} for final state showering and multi-particle
interactions. Background events are all generated by PYTHIA.

When generating the signal and background samples we include effects
of the initial state radiation (ISR) but assume that accelerator beams
are mono-energetic.
We neglect beamstrahlung due to beam-beam interactions, which result
in the additional energy smearing and increase the fraction of $e^+
e^-$ pairs colliding with lower energies, as it depends strongly on
the accelerator design and assumed beam parameters.
Using ILC beam spectra modeled with CIRCE1~\cite{circe} we estimated
that the influence of beamstrahlung on the number of expected signal
events was of the order of 1--3\%. For the background processes increase
of up to 10\% in the event rate was observed.

Both signal and background event rates depend also on the expected
polarization of the electron and positron beams. By a proper choice of the beam polarization one can significantly improve the signal to background ratio. However, possible degree of the beam polarization is determined by the accelerator design. Therefore, for the sake of generality, we consider unpolarized $e^+e^-$ beams.

The jet reconstruction is performed by FASTJET 2.4.1 \cite{fastjet1,fastjet2}
using anti-kt algorithm with a cone size of 0.4.
To take into account the detector effects, we include acceptance cuts
and simple modeling of the jet energy resolution \cite{Zarnecki:Top2015}:
\beq
\frac{\sigma_E}{E} = \begin{cases} \frac{\cal S}{\sqrt{E_j[\text{GeV}]}} &
  \mbox{for } E_j < 100 \text{ GeV},\\
~~~\frac{\cal S}{\sqrt{100}} & \mbox{for } E_j \geq 100 \text{ GeV}, \end{cases}
\label{smear}
\eeq
where $\frac{\sigma_E}{E}$ is the relative jet energy uncertainty
which is used to smear the measured jet energy and $\cal S$ is
the resolution parameter.
The formula \eqref{smear} describes well the energy resolution expected
for single high energy jets when using algorithms based on
so called particle flow~\cite{Thomson:2009rp}.
Test data analysis and results of the detailed detector simulations
based on GEANT4 indicate that the relative energy resolution of 3--4\%
should be feasible for highest jet energies both at ILC and CLIC,
corresponding to parameter $\cal S$=30--40\%.
However, the measurement precision is expected to deteriorate when we
take into account the influence of beam related backgrounds and
effects related to detector acceptance or event reconstruction.
Therefore, we take a conservative value of $\cal S$=50\% for our study.
Final state leptons (electrons and muons) are assumed to be reconstructed
without any sizable uncertainty, as determination of their momentum
will be based on the track measurement.
After signal and background events are generated
and reconstructed, the analysis is carried out using ROOT 5.34
\cite{root}.


\section{$e^+ e^- \rightarrow H^+ H^-$}
\label{eeHH}

\begin{figure}[t]
\begin{tikzpicture}[node distance=1.5cm,thick, rounded corners=0pt,line cap=round]
\begin{scope}[xshift=0cm]
\coordinate[] (v1);
\coordinate[right=of v1] (v2);
\coordinate[above right=of v2] (v3);
\coordinate[below right=of v2] (v4);
\coordinate[below left=of v1] (i1);
\coordinate[above left=of v1] (i2);
\coordinate[position=-30 degrees from v3] (o1);
\coordinate[position=30 degrees from v3] (v5);
\coordinate[position=30 degrees from v4] (o2);
\coordinate[position=-30 degrees from v4] (v6);
\coordinate[position=15 degrees from v5] (mu);
\coordinate[position=-15 degrees from v5] (nu);
\coordinate[position=15 degrees from v6] (j1);
\coordinate[position=-15 degrees from v6] (j2);
\draw[fermion] (i1)node[left]{$e^-$} -- (v1);
\draw[fermion](v1)--(i2)node[left]{$e^+$};
\draw[boson] (v1) --node[below]{$Z,\gamma$} (v2);
\draw[cscalar] (v3) --node[above left]{$H^+$} (v2);
\draw[cscalar] (v2) --node[below left]{$H^-$} (v4);
\draw[boson](v3)--node[above]{$W^+$}(v5);
\draw[scalar](v3)--(o1)node[right]{$H$};
\draw[boson](v4)--node[below]{$W^-$}(v6);
\draw[scalar](o2)node[right]{$H$}--(v4);
\draw[fermion](nu)node[right]{$\nu$}--(v5);
\draw[fermion](v5)--(mu)node[right]{$\lep$};
\draw[jet](v6)--(j2)node[right]{$j$};
\draw[jet](j1)node[right]{$j$}--(v6);
\filldraw [black] (v1) circle (2pt);
\filldraw [black] (v2) circle (2pt);
\filldraw [black] (v3) circle (2pt);
\filldraw [black] (v4) circle (2pt);
\filldraw [black] (v5) circle (2pt);
\filldraw [black] (v6) circle (2pt);
\end{scope}
\begin{scope}[xshift=8cm]
\coordinate[] (v1);
\coordinate[right=of v1] (v2);
\coordinate[above right=of v2] (v3);
\coordinate[below right=of v2] (v4);
\coordinate[below left=of v1] (i1);
\coordinate[above left=of v1] (i2);
\coordinate[position=-30 degrees from v3] (o1);
\coordinate[position=30 degrees from v3] (v5);
\coordinate[position=30 degrees from v4] (o2);
\coordinate[position=-30 degrees from v4] (v6);
\coordinate[position=15 degrees from v5] (mu);
\coordinate[position=-15 degrees from v5] (nu);
\coordinate[position=15 degrees from v6] (j1);
\coordinate[position=-15 degrees from v6] (j2);
\draw[fermion] (i1)node[left]{$e^-$} -- (v1);
\draw[fermion](v1)--(i2)node[left]{$e^+$};
\draw[boson] (v1) --node[below]{$Z,\gamma$} (v2);
\draw[cscalar] (v3) --node[above left]{$H^+$} (v2);
\draw[cscalar] (v2) --node[below left]{$H^-$} (v4);
\draw[boson](v3)--node[above]{$W^+$}(v5);
\draw[scalar](v3)--(o1)node[right]{$H$};
\draw[boson](v4)--node[below]{$W^-$}(v6);
\draw[scalar](o2)node[right]{$H$}--(v4);
\draw[jet](nu)node[right]{$j$}--(v5);
\draw[jet](v5)--(mu)node[right]{$j$};
\draw[jet](v6)--(j2)node[right]{$j$};
\draw[jet](j1)node[right]{$j$}--(v6);
\filldraw [black] (v1) circle (2pt);
\filldraw [black] (v2) circle (2pt);
\filldraw [black] (v3) circle (2pt);
\filldraw [black] (v4) circle (2pt);
\filldraw [black] (v5) circle (2pt);
\filldraw [black] (v6) circle (2pt);
\end{scope}
\\
\vspace{1cm}
\begin{scope}[xshift=4cm, yshift=-5cm]
\coordinate[] (v1);
\coordinate[right=of v1] (v2);
\coordinate[above right=of v2] (v3);
\coordinate[below right=of v2] (v4);
\coordinate[below left=of v1] (i1);
\coordinate[above left=of v1] (i2);
\coordinate[position=-30 degrees from v3] (o1);
\coordinate[position=30 degrees from v3] (v5);
\coordinate[position=30 degrees from v4] (o2);
\coordinate[position=-30 degrees from v4] (v6);
\coordinate[position=15 degrees from v5] (mu);
\coordinate[position=-15 degrees from v5] (nu);
\coordinate[position=15 degrees from v6] (j1);
\coordinate[position=-15 degrees from v6] (j2);
\draw[fermion] (i1)node[left]{$e^-$} -- (v1);
\draw[fermion](v1)--(i2)node[left]{$e^+$};
\draw[boson] (v1) --node[below]{$Z,\gamma$} (v2);
\draw[cscalar] (v3) --node[above left]{$H^+$} (v2);
\draw[cscalar] (v2) --node[below left]{$H^-$} (v4);
\draw[boson](v3)--node[above]{$W^+$}(v5);
\draw[scalar](v3)--(o1)node[right]{$H$};
\draw[boson](v4)--node[below]{$W^-$}(v6);
\draw[scalar](o2)node[right]{$H$}--(v4);
\draw[fermion](nu)node[right]{$\nu$}--(v5);
\draw[fermion](v5)--(mu)node[right]{$\lep$};
\draw[fermion](v6)--(j2)node[right]{$\lep$};
\draw[fermion](j1)node[right]{$\nu$}--(v6);
\filldraw [black] (v1) circle (2pt);
\filldraw [black] (v2) circle (2pt);
\filldraw [black] (v3) circle (2pt);
\filldraw [black] (v4) circle (2pt);
\filldraw [black] (v5) circle (2pt);
\filldraw [black] (v6) circle (2pt);
\end{scope}

\end{tikzpicture}
\caption{The Feynman diagrams for charged scalar pair production and
  decay processes, $e^+e^-\to ( H^+H^-\to W^+W^-HH\to ) \lep\nu jj HH$,
  $e^+e^-\to jj jj HH$ and $e^+e^-\to \lep\nu \lep \nu HH$.}
\label{fig_feynman}
\end{figure}
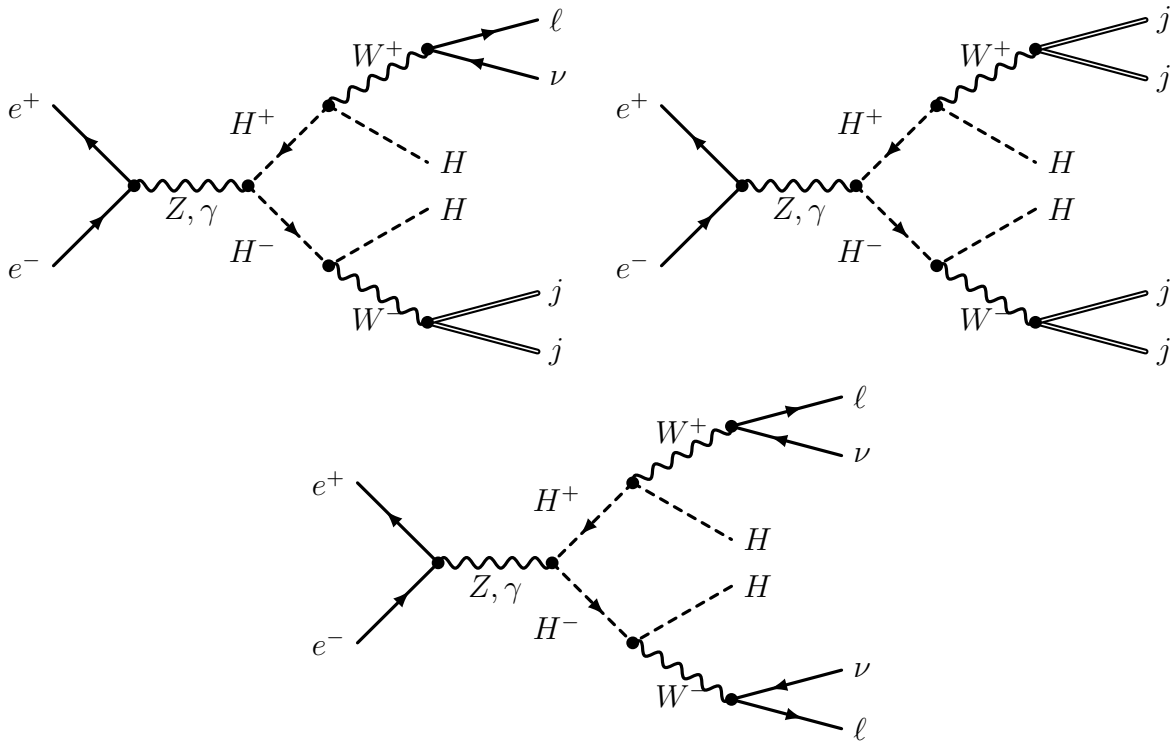

In this section, the analysis of  the charged scalar pair production
process is considered for different benchmark
scenarios and different final states.
Two final states are considered for signal events
($\lep\nu jj HH $ and $jjjj HH $) and for each final state an
independent analysis based on kinematic selection cuts is
performed.
The third final state, $\lep\nu\lep\nu HH $, is much more
difficult as the two leptons come from different $W^\pm$ and it
is not possible to make use of kinematic constraints for efficient
background suppression.
Still, if the signal is observed in other channels, a dedicated analysis
could be performed, based on single lepton energy spectra.
In \cref{fig_feynman}, diagrams for  charged scalar pair production
with different decay channels are shown.
In what follows, cross sections of signal and
background processes are presented for two selected decay channels.
Then, event generation and
analysis are described in detail including selection cuts and their
efficiencies.

\subsection{Signal and Background Cross Sections}

Cross sections of signal event production are calculated at leading
order (LO) using a Monte
Carlo simulation performed by CompHEP. For decay branching ratios of
Standard Model particles like $ W$ and $Z$ bosons, standard PDG values \cite{pdg} are used.
The charged scalar branching ratios are taken from 2HDMC
1.6.3 \cite{2hdmc1,2hdmc2}. Cross sections of background processes are
calculated with PYTHIA.
We consider the following background processes: $W^+ W^-$ pair
production (WW), $Z Z$ pair production (ZZ), fermion pair production
from $e+ e-$ annihilation into single $Z^\star / \gamma^\star $ (Z)
and the top pair production (TT).
\Cref{xsec500gev,xsec1tev}
show the LO signal and background cross sections at center of mass
energies of 0.5 and 1 TeV, respectively.

\begin{table}[b]
\centering
\begin{tabular}{||c|c|c|c|c|c|c|c||}
\hline
\multirow{2}{*}{Process} & \multicolumn{3}{c|}{$H^+H^-$} &
\multicolumn{4}{c||}{Background processes} \\ \cline{2-8}
& BP1 & BP2 & BP3 &  WW  & ZZ  & Z & TT \\
\hline\hline
Cross section [fb] & 82.2 &  70.9 &  44.6 & 7807 & 583 & 16790 & 595 \\
\hline
\end{tabular}
\caption{Signal and background cross sections at $\sqrt{s}=0.5$ TeV. \label{xsec500gev}}
\end{table}

\begin{table}[t]
\centering
\begin{tabular}{||c|c|c|c|c|c|c|c||}
\hline
\multirow{2}{*}{Process} & \multicolumn{3}{c|}{$H^+H^-$} &
\multicolumn{4}{c||}{Background processes} \\ \cline{2-8}
& BP1 & BP2 & BP3 &  WW  & ZZ  & Z & TT \\
\hline\hline
Cross section [fb] & 28.1 & 27.3 & 25.3 & 3180 & 233 & 4304 & 212 \\
\hline
\end{tabular}
\caption{Signal and background cross sections at $\sqrt{s}=1$ TeV. \label{xsec1tev}}
\end{table}

\subsection{Event Generation and Analysis}
\label{evgenana}

The charged scalar pair production, with subsequent decay $H^{\pm} \ra
W^{\pm}H$, where $W$ boson decays either to a muon-neutrino or two light
jet pairs,  is generated using CompHep.
 Depending on the $W$ boson decay channel, three final states are
produced, i.e., $\lep\nu\lep\nu HH$, $\lep\nu jjHH$ and $jjjjHH$. These
final states are labeled fully leptonic, semi-leptonic and fully
hadronic, respectively.
As mentioned above, we consider semi-leptonic and fully
hadronic channels only, and an independent analysis based on kinematic
selection cuts is performed for each final state.
In what follows, the analysis and selection cuts of different
final states are described in detail.

\subsubsection{Semi-leptonic Final State}
\label{sec:lorentz}

Signal events are characterized by a single lepton and two light jets
from $W$ boson decays, and missing transverse momentum.
Therefore, we require to have one lepton and two jets passing a 10 GeV
transverse energy threshold reconstructed in the event.
The threshold cut is applied to reject events with soft leptons or
jets in their final state and the same cut is applied at both center
of mass energies, $\sqrt{s}=0.5$ and 1 TeV.
Although a harder cut could be adopted at 1 TeV collisions,
the current set of selection cuts gives a reasonable background suppression while limiting the loss of statistics in signal selection.
The jet energy threshold is set to avoid uncertainties in
soft jet energy measurement due to the jet reconstruction
algorithm and detector effects.
The threshold of missing transverse momentum is
taken to be 20 GeV for the two center of mass energies. This cut is
most useful for suppression of single or pair production of $Z$ bosons.

\begin{figure}[tb]
\centering
\begin{subfigure}{0.5\textwidth}
  \centering
  \includegraphics[width=\textwidth]{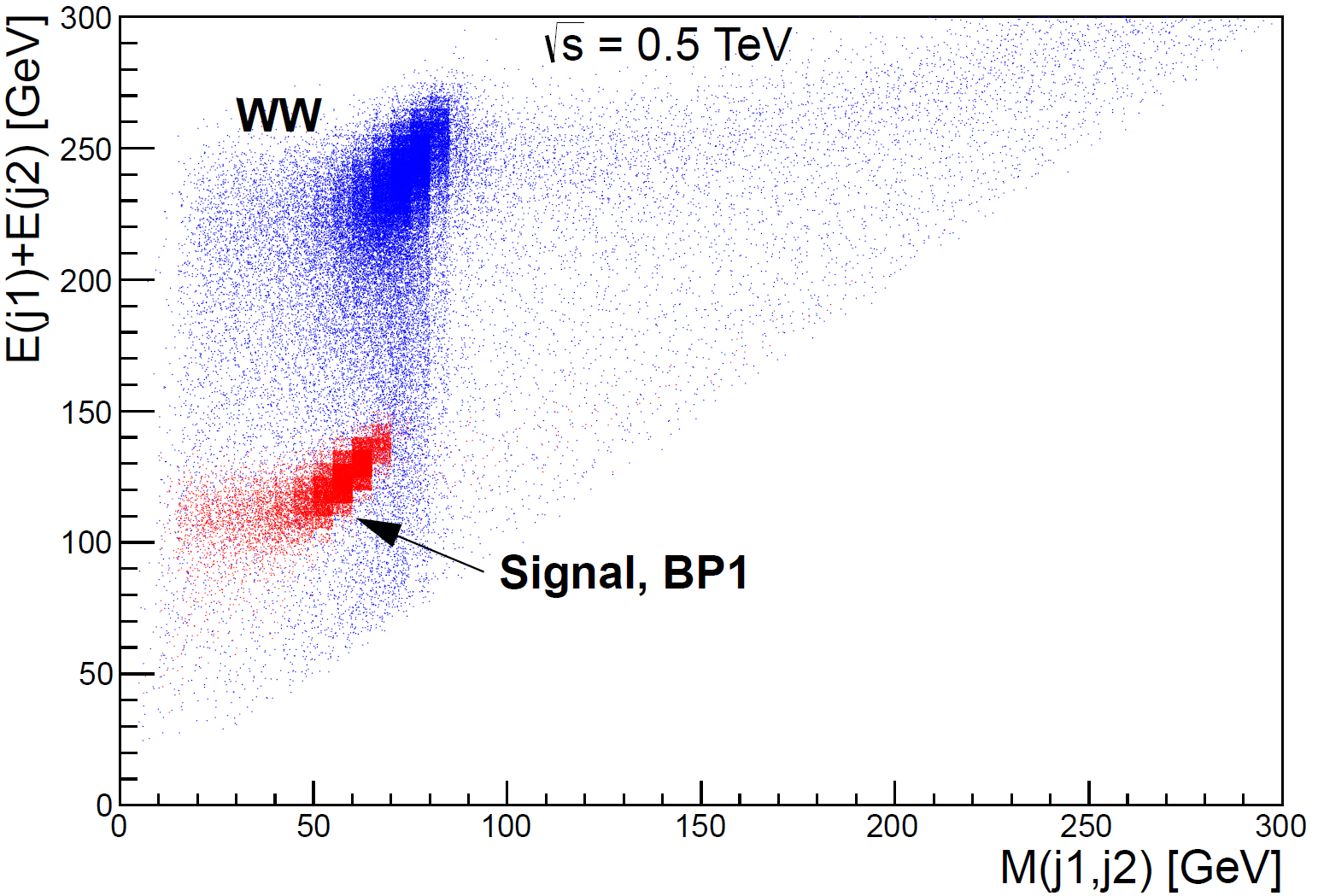}
  \caption{$\sqrt{s}=0.5$ TeV}
  \label{2D}
\end{subfigure}%
\begin{subfigure}{0.5\textwidth}
  \centering
  \includegraphics[width=\textwidth]{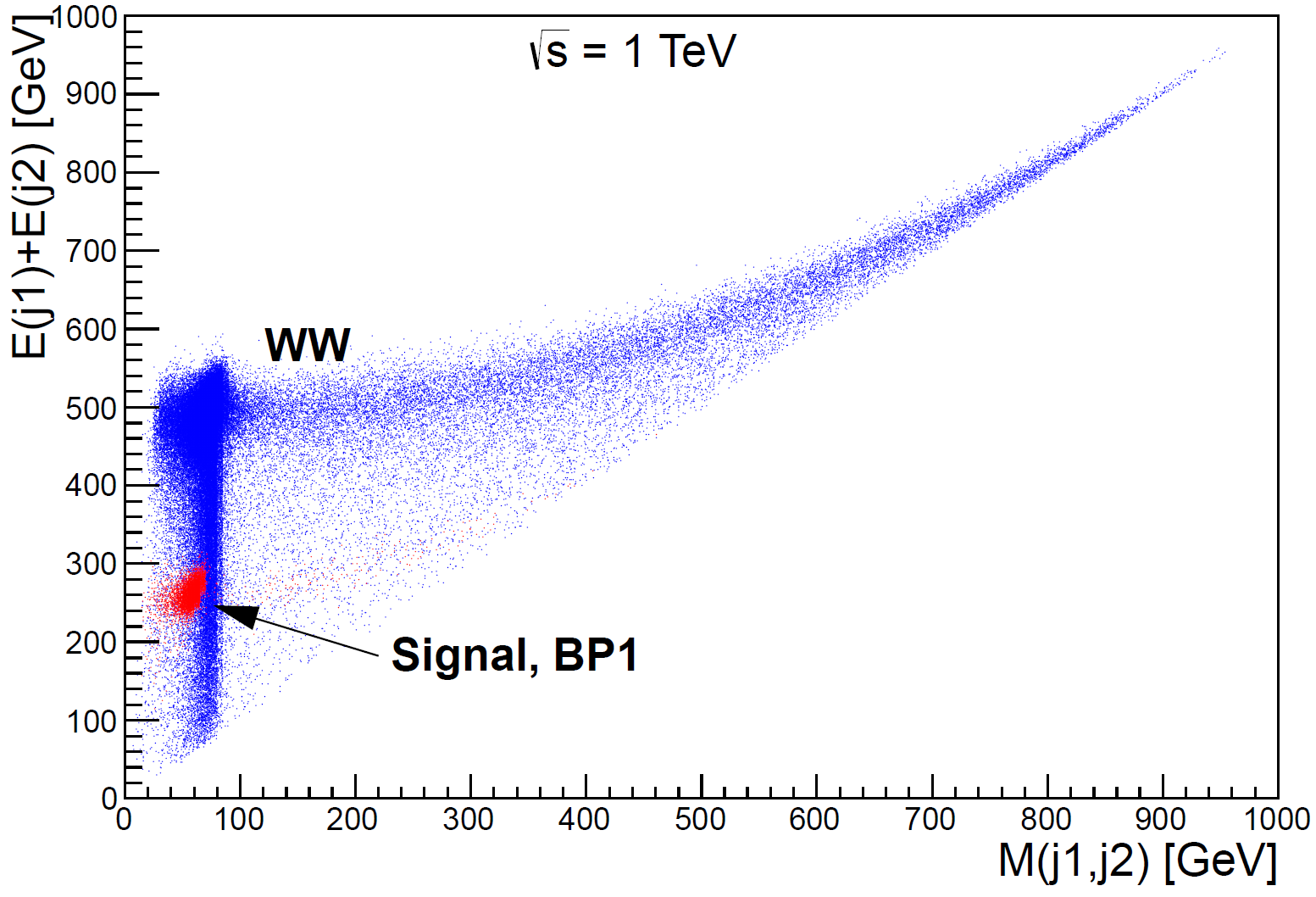}
  \caption{$\sqrt{s}=1$ TeV}
  \label{2D2}
\end{subfigure}
\caption{Correlation for $e^+e^-\to H^+H^-$ between the sum of energies of two jets and their invariant mass at $\sqrt{s}=0.5$ TeV (left) and 1 TeV (right). The comparison is only between the semi-leptonic signal (BP1) and WW background.}
\label{}
\end{figure}

For all benchmark scenarios considered in our analysis  the scalar
mass difference  $m_{H^\pm}-m_{H}$  is significantly smaller than the
nominal $W^\pm$ boson mass.
For signal events, two jets observed in semi-leptonic channel
come from the decay of the off-shell $W^\pm$ boson ($W^\star$), with two jet
invariant mass corresponding to the  $W^\star$ virtuality.
On the other hand, the sum of jet energies measured in the laboratory
frame is given by a product of the the $W^\star$ virtuality (i.e. the
energy in the $W^\star$ rest frame) and the Lorenz boost factor
$\gamma$ corresponding to the $W^\star$ velocity in the laboratory frame.
The Lorentz boost of $W^\star$ is unknown, but we expect that
$W^\star$ production with the highest possible virtuality is most likely.
In such a case, $W^\star$ is almost at rest in the reference frame of
decaying $H^{\pm}$, and we can use the Lorentz boost factor of charged
scalar for transformation of jet energies.
As the energies of the charged scalars are
given by the beam energy, their Lorentz boost factor is
uniquely defined by their mass: $\gamma = {E_{beam}}/{M_{H^\pm}}$.
Therefore we expect to observe a narrow peak not only for the two
jet invariant mass but also in the sum of two jet energies distribution.
This is clearly seen in \cref{2D,2D2}, where
the sum of two jet energies is plotted versus their invariant mass,
for benchmark scenario BP1 (red dots).
For comparison, semi-leptonic background events $e^+e^-\to  W^+W^-$,
dominated by on-shell $W^\pm$ boson decays, are also shown.
The peak observed for signal events is clearly shifted with respect to the background event
distribution, both in the energy
and in the invariant mass, and this observation can be used for efficient
suppression of background events.

It should be mentioned that background events in the semi-leptonic
channel can also be suppressed by a cut on the missing mass , i.e. on
invariant mass of the final state particles escaping detection,
reconstructed  from the energy-momentum
conservation~\cite{Ginzburg:2014ora}.
For signal events the missing mass is at least twice the $H$ boson mass,
whereas for $e^+e^-\to  W^+W^-$ events (in semi-leptonic decay channel) its
distribution should be peaked at small masses, corresponding to
a single escaping neutrino.
However, the cut on the missing mass is strongly correlated with the
jet energy and invariant mass cuts and therefore the resulting improvement in
the event selection is marginal. We do not use this cut for the
presented results.

\begin{figure}[tb]
\centering
\begin{subfigure}{0.5\textwidth}
  \centering
  \includegraphics[width=\textwidth]{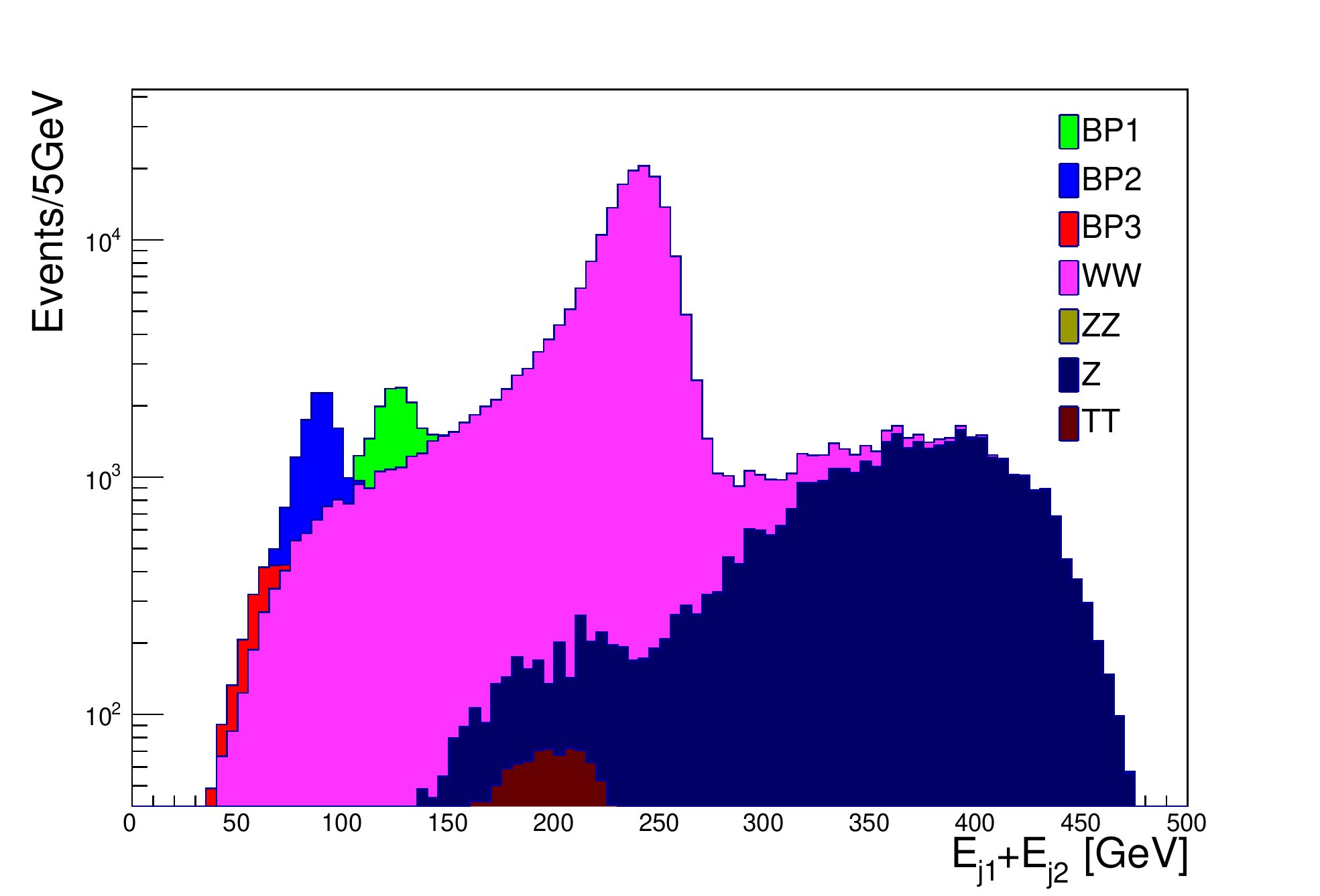}
  \caption{$e^{+}e^{-}\ra \ell\nu jjHH$ at $\sqrt{s}=0.5$ TeV}
  \label{ej1j2_lnujjHH}
\end{subfigure}%
\begin{subfigure}{0.5\textwidth}
  \centering
  \includegraphics[width=\textwidth]{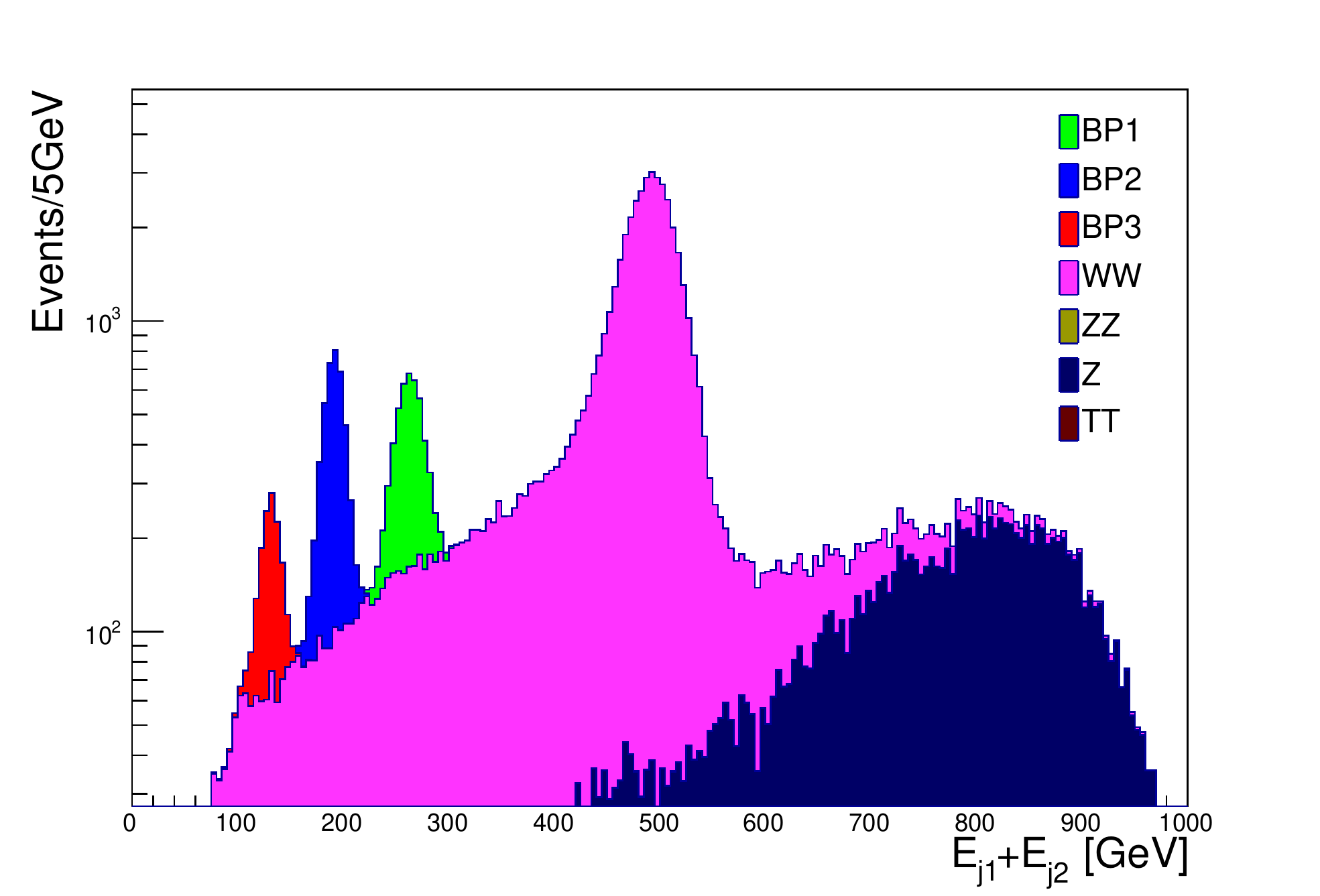}
  \caption{$e^{+}e^{-}\ra \ell\nu jjHH$ at $\sqrt{s}=1$ TeV}
  \label{ej1j2_1_lnujjHH}
\end{subfigure}
\caption{Sum of two jets energies in semileptonic final state at $\sqrt{s}=0.5$ TeV (left) and 1 TeV (right), for $e^+e^-\to H^+H^-$.
}
\label{}
\end{figure}

We found that the jet energy sum distribution itself, as shown in
\cref{ej1j2_lnujjHH,ej1j2_1_lnujjHH} gives good signal
and background separation.
As described above, the distribution is softer for
signal events due to the smaller $W^\pm$ boson virtuality and smaller
Lorentz boost factor.
For background events, like $WW$ and $ZZ$ pair production, the two-jet
energy is the energy of the parent boson, which is half of the
collision energy.

Based on \cref{2D,2D2}, a cut on the sum of jet
energies is applied for semi-leptonic events, of 150(350)  GeV for
$\sqrt{s}=0.5$(1) TeV center of mass energy. \Cref{selcut1} gives a
summary of selection cuts for this final state
while \cref{cuteff1,cuteff2} present selection
efficiencies for signal and background processes at $\sqrt{s}=0.5$ and 1 TeV,
respectively.

\begin{figure}[htb]
\centering
\begin{subfigure}{0.5\textwidth}
  \centering
  \includegraphics[width=\textwidth]{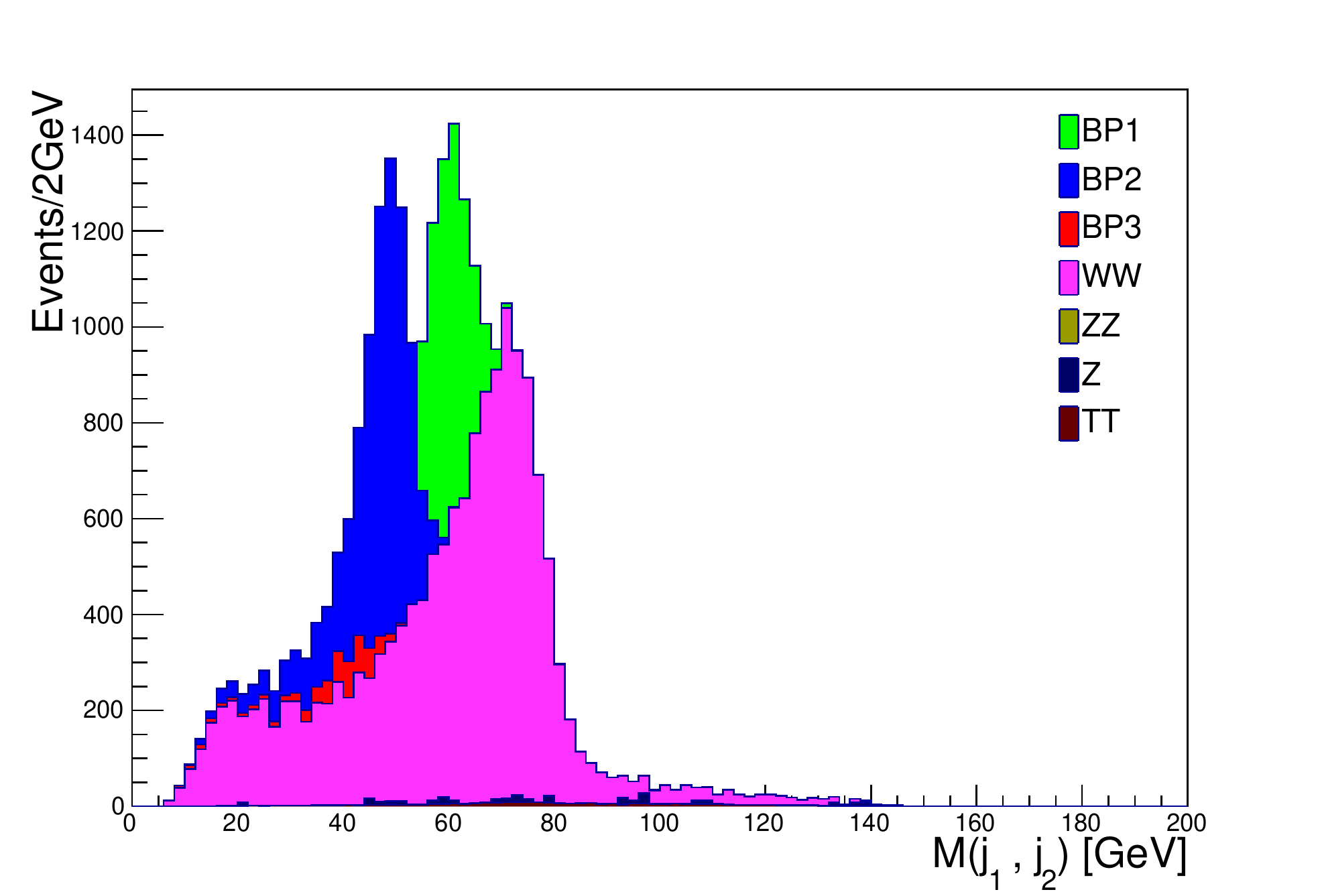}
  \caption{$e^{+}e^{-}\ra \ell\nu jjHH$ at $\sqrt{s}=0.5$ TeV}
  \label{mj1j2_lnujjHH}
\end{subfigure}%
\begin{subfigure}{0.5\textwidth}
  \centering
  \includegraphics[width=\textwidth]{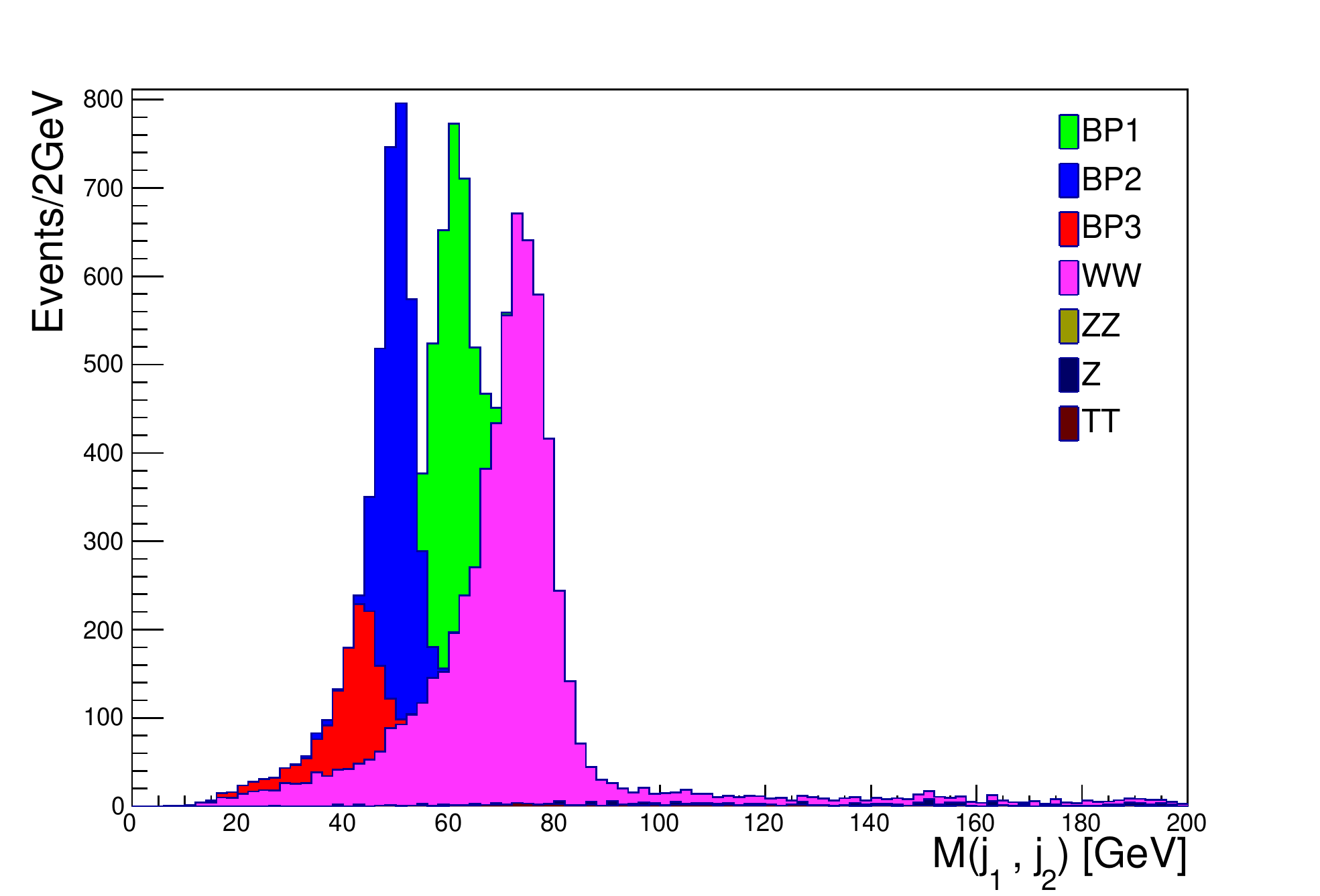}
  \caption{$e^{+}e^{-}\ra \ell\nu jjHH$ at $\sqrt{s}=1$ TeV}
  \label{mj1j2_1_lnujjHH}
\end{subfigure}
\caption{Invariant mass of two jets in semileptonic final state at $\sqrt{s}=0.5$ TeV (left) and 1 TeV (right), for $e^+e^-\to H^+H^-$.}
\label{}
\end{figure}

Invariant mass distribution of the two jets,
obtained after applying the cut on the sum of the two jet energies,
is plotted in \cref{mj1j2_lnujjHH} (
\ref{mj1j2_1_lnujjHH}) for signal and background
events at $\sqrt s=0.5$ TeV (1 TeV).
These plots are used to obtain information
about the mass difference $m_{H^{\pm}}-m_H$ with statistical
uncertainty below 100~MeV (see discussion in \cref{sec:dm_mass}).
Using a mass window cut  the final numbers of signal (S) and
background (B) events can be extracted, as well as the signal significance
 $s = S/\sqrt{S+B}$,  as shown in  \cref{lnujjHH_finaleff}.
Significance of the signal observation is very high even for
the least favoured benchmark point, BP3.
The corresponding precision of the signal cross section determination
is 2--12\% for $\sqrt s=0.5$ TeV and 2--4\% for $\sqrt s=1$ TeV.

\begin{table}[htb]
\centering
\begin{tabular}{|c|c|c|}
\hline
\hline
\multicolumn{3}{|c|}{$H^{+}H^{-}$ analysis, semi-leptonic final state selection}\\
\hline
Selection cut & $\sqrt{s}=0.5$ TeV & $\sqrt{s}=1$ TeV\\
\hline
One lepton & $E_{T}>10$ GeV &  $E_{T}>10$ GeV \\
\hline
Two jets & $E_{T}>10$ GeV &  $E_{T}>10$ GeV \\
\hline
$\met$ & $\met ~>~ 20$ GeV & $\met~>~ 20$ GeV \\
\hline
$E(j_{1})+E(j_{2})$ & $E(j_{1})+E(j_{2})~<~150$ GeV & $E(j_{1})+E(j_{2})~<~350$ GeV\\
\hline
\hline
\end{tabular}
\caption{Selection cuts for semi-leptonic final state analysis at two center of mass energies of 0.5 and 1 TeV.\label{selcut1}}
\end{table}

\begin{table}[htb]
\centering
\begin{tabular}{|c|c|c|c|c|c|c|c|}
\hline
\hline
\multicolumn{8}{|c|}{$H^{+}H^{-}$ analysis, semi-leptonic final state selection}\\
\hline
Cut eff. & BP1 & BP2 & BP3 & WW & ZZ & Z & TT \\
\hline
One Lepton & 0.89 & 0.93 & 0.77 & 0.45 & 0.13 & 0.069 & 0.67\\
\hline
Two Jets & 0.67 & 0.78 & 0.53 & 0.49 & 0.23 & 0.25 & 0.027\\
\hline
$\met$ & 0.83 & 0.88 & 0.49 & 0.78 & 0.061 & 0.35 & 0.92\\
\hline
$E(j_1)+E(j_2)$ & 1 & 1 & 1 & 0.08 & 0.12 & 0.0065 & 0.18\\
\hline
Total eff. & 0.5 & 0.64 & 0.2 & 0.014 & 0.00021 & 3.9e-05 & 0.0029\\
\hline
\hline
\end{tabular}
\caption{Cut efficiencies for semi-leptonic final state analysis at center of mass energy of 0.5 TeV.\label{cuteff1}}
\end{table}

\begin{table}[htb]
\centering
\begin{tabular}{|c|c|c|c|c|c|c|c|}
\hline
\hline
\multicolumn{8}{|c|}{$H^{+}H^{-}$ analysis, semi-leptonic final state selection}\\
\hline
Cut eff. & BP1 & BP2 & BP3 & WW & ZZ & Z & TT \\
\hline
One Lepton & 0.95 & 0.97 & 0.95 & 0.28 & 0.11 & 0.063 & 0.57\\
\hline
Two Jets & 0.87 & 0.91 & 0.68 & 0.47 & 0.21 & 0.23 & 0.024\\
\hline
$\met$ & 0.97 & 0.97 & 0.9 & 0.86 & 0.14 & 0.51 & 0.93\\
\hline
$E(j_1)+E(j_2)$ & 1 & 1 & 1 & 0.12 & 0.11 & 0.013 & 0.25\\
\hline
Total eff. & 0.8 & 0.86 & 0.59 & 0.014 & 0.00035 & 9.6e-05 & 0.0032\\
\hline
\hline
\end{tabular}
\caption{Cut efficiencies for semi-leptonic final state analysis at center of mass energy of 1 TeV.\label{cuteff2}}
\end{table}
\begin{table}[t]
\centering
\begin{tabular}{||c|c|c|c|c||c|c|c|c||}
\hline\hline
\multicolumn{9}{|c|}{$H^{+}H^{-}$, semi-leptonic final state at ${\cal L}=500$~fb$^{-1}$}\\
\hline
& \multicolumn{4}{c||}{$\sqrt{s}=$0.5 TeV}& \multicolumn{4}{c|}{$\sqrt{s}=$1 TeV}\\
\hline
& S& B& S/B & $S/\sqrt{S+B}$ & S & B & S/B & $S/\sqrt{S+B}$\\
\hline\hline
BP 1 & 5101 & 6136 & 0.83 & 48  & 3055 & 1901 & 1.6 & 43\\
BP 2 & 4885 & 2285 & 2.1  & 58  & 2590 & 461  & 5.6 & 47 \\
BP 3 & 474  & 2784 & 0.17 & 8.3 & 861  & 433  & 2.0 & 24 \\
\hline
\end{tabular}
\caption{Number of events in signal and background processes after all selection cuts at integrated luminosity of 500~fb$^{-1}$.  S and B stand for the number of signal and background events. \label{lnujjHH_finaleff}}
\end{table}

\subsubsection{Fully Hadronic Final State}

\begin{figure}[tb]
\centering
\begin{subfigure}{0.5\textwidth}
  \centering
  \includegraphics[width=\textwidth]{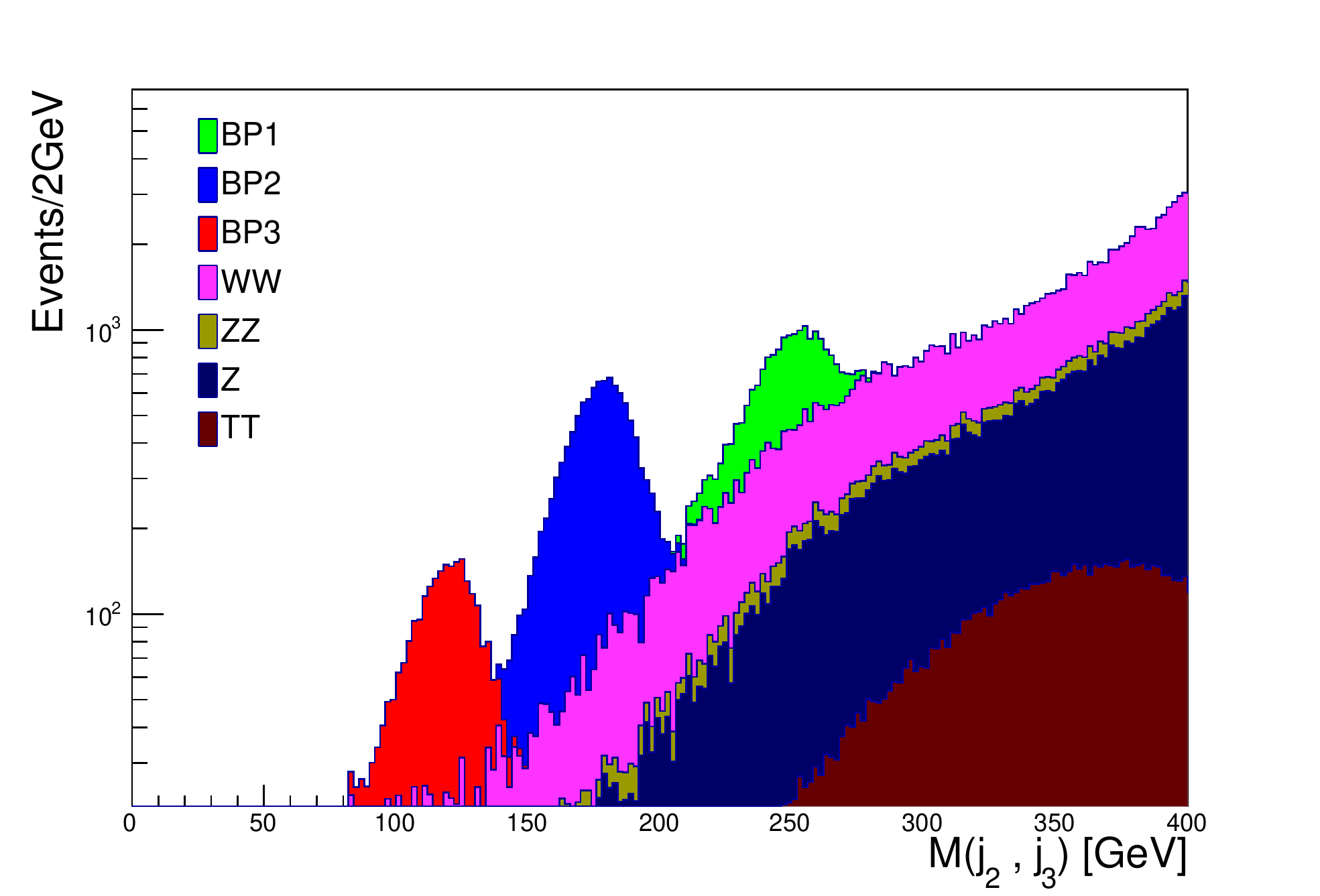}
  \caption{$e^{+}e^{-}\ra jjjjHH$ at $\sqrt{s}=0.5$ TeV}
  \label{ej1j2j3j4_jjjjHH}
\end{subfigure}%
\begin{subfigure}{0.5\textwidth}
  \centering
  \includegraphics[width=\textwidth]{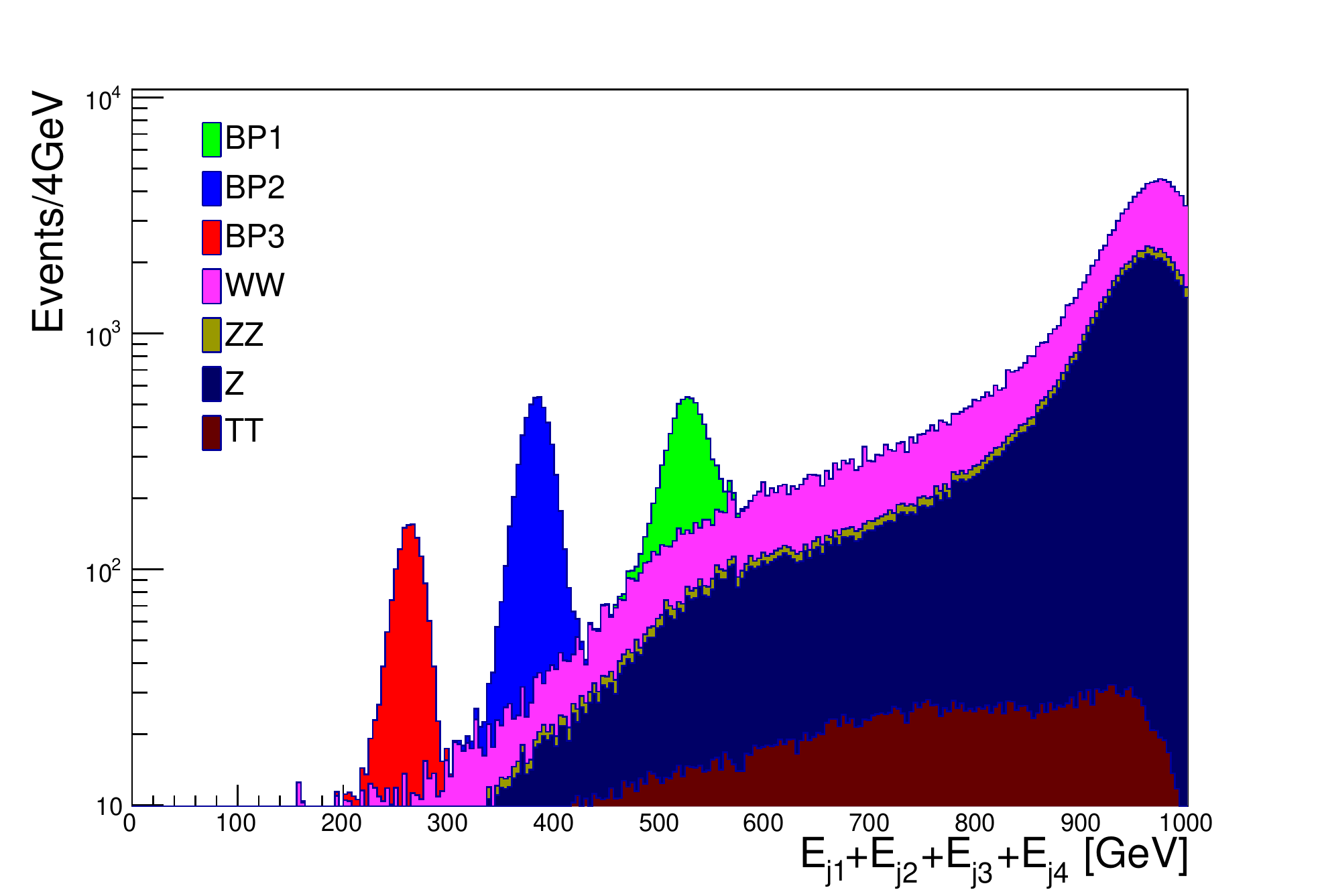}
  \caption{$e^{+}e^{-}\ra jjjjHH$ at $\sqrt{s}=1$ TeV}
  \label{ej1j2j3j4_1_jjjjHH}
\end{subfigure}
\caption{Sum of four jets energies in fully hadronic final state at $\sqrt{s}=0.5$ TeV (left) and 1 TeV (right), for $e^+e^-\to H^+H^-$.}
\label{}
\end{figure}

Signal events are characterized by four light jets from off-shell $W$ bosons
decays and missing transverse momentum. An event is required to have four jets
passing 10 GeV transverse energy threshold, both for $\sqrt{s}=$ 0.5 and
1 TeV.
Contrary to the case of semi-leptonic final state, there is no
$\met$ within the detector resolution due to the back-to-back configuration
of $H$ bosons which results in cancellation of their effect in $\met$
calculation.
Therefore the missing transverse momentum is required to be
less than 10 GeV.
The sum of energies of the four jets in the event is
required to be less than 300(600) GeV at $\sqrt{s}=0.5$(1) TeV
respectively.
This cut is applied following the same strategy as
described in the semi-leptonic case using a two dimensional
correlation plot.
The distributions of the sum of energies of the four
jets are shown in \cref{ej1j2j3j4_jjjjHH,ej1j2j3j4_1_jjjjHH},
where in case of background events like $WW$
and $ZZ$, the four jet energy is consistent with the collision
energy and is expected to dominate the region near 0.5
or 1 TeV, depending on center of mass energy. \Cref{selcut2}
summarizes selection cuts  while \cref{cuteff3,cuteff4} present
selection efficiencies for signal and background
processes at $\sqrt{s}=0.5$ and 1 TeV, respectively.

\begin{table}[h]
\centering
\begin{tabular}{|c|c|c|}
\hline
\hline
\multicolumn{3}{|c|}{$H^{+}H^{-}$ analysis, fully hadronic final state selection}\\
\hline
Selection cut & $\sqrt{s}=0.5$ TeV & $\sqrt{s}=1$ TeV\\
\hline
4 jets & $E_{T}>10$ GeV &  $E_{T}>10$ GeV \\
\hline
$\met$ & $\met~<~10$ GeV & $\met~<~10$ GeV \\
\hline
$\sum\limits_{i=1}^{4}E(j_i)$ & $\sum\limits_{i=1}^{4}E(j_i)~<~300$ GeV & $\sum\limits_{i=1}^{4}E(j_i)~<~600$ GeV\\
\hline
\hline
\end{tabular}
\caption{Selection cuts for fully hadronic final state analysis at two center of mass energies of 0.5 and 1 TeV.\label{selcut2}}
\end{table}

\begin{table}[h]
\centering
\begin{tabular}{|c|c|c|c|c|c|c|c|}
\hline
\hline
\multicolumn{8}{|c|}{$H^{+}H^{-}$ analysis, fully hadronic final state selection}\\
\hline
Cut eff. & BP1 & BP2 & BP3 & WW & ZZ & Z & TT \\
\hline
Four Jets & 0.48 & 0.57 & 0.42 & 0.19 & 0.23 & 0.049 & 0.1\\
\hline
$\met$ &1 & 1 & 1 & 0.94 & 0.83 & 0.88 & 0.69\\
\hline
$\sum\limits_{i=1}^{4}E(j_i)$&1 & 1 & 1 & 0.053 & 0.065 & 0.036&0.25 \\ \hline
Total eff. & 0.48 & 0.57 & 0.42 & 0.0095 & 0.012 & 0.0016 & 0.018\\
\hline
\hline
\end{tabular}
\caption{Cut efficiencies for fully hadronic final state analysis at center of mass energy of 0.5 TeV.\label{cuteff3}}
\end{table}

\begin{table}[h]
\centering
\begin{tabular}{|c|c|c|c|c|c|c|c|}
\hline
\hline
\multicolumn{8}{|c|}{$H^{+}H^{-}$ analysis, fully hadronic final state selection}\\
\hline
Cut eff. & BP1 & BP2 & BP3 & WW & ZZ & Z & TT \\
\hline
Four Jets & 0.77 & 0.81 & 0.54 & 0.11 & 0.15 & 0.055 & 0.11\\
\hline
$\met$ &0.86 & 0.93 & 0.9 & 0.91 & 0.76 & 0.84 & 0.61\\
\hline
$\sum\limits_{i=1}^{4}E(j_i)$ &1&1 & 1 & 0.045 & 0.051 & 0.037 & 0.25\\
\hline
Total eff. & 0.66 & 0.75 & 0.48 & 0.0045 & 0.0058 & 0.0017 & 0.017\\
\hline
\hline
\end{tabular}
\caption{Cut efficiencies for fully hadronic final state analysis at center of mass energy of 1 TeV.\label{cuteff4}}
\end{table}

When the cut on the sum of four jets energies is applied, the
invariant mass of pairs of jets can be investigated.
As described in \cref{sec:lorentz}, both $W^\star$ bosons are likely to
be produced with the same virtuality and with the same Lorentz boost
factor as for  $H^\pm$.
As the sum of energies for both jet pairs is expected to be the same,
the jet with the highest energy in the laboratory reference frame
should be matched to the jet with the lowest energy.
Consequently, the second and third jet should be
matched to reconstruct the other $W$ boson in the event.
The described matching is not fully efficient as
detector effects can disturb the jet energy ordering.
More detailed analysis, based on the so called kinematic
fit\footnote{In the kinematic fit procedure all possible jet combinations
  are considered. Each hypothesis is compared with the expected event
  topology and kinematic constraints, taking into account the detector
  resolution, acceptance and reconstruction efficiency,
  based on the detailed simulation of detector effect.
  The likelihood value is calculated for each combination and the
  hypothesis with the highest  likelihood is selected for the
  analysis as the proper one.}
approach, would allow to select the correct (most probable) jet
matching with higher efficiency.
%
This is beyond the scope of present analysis.

\begin{figure}
\centering
\begin{subfigure}{0.5\textwidth}
  \centering
  \includegraphics[width=\textwidth]{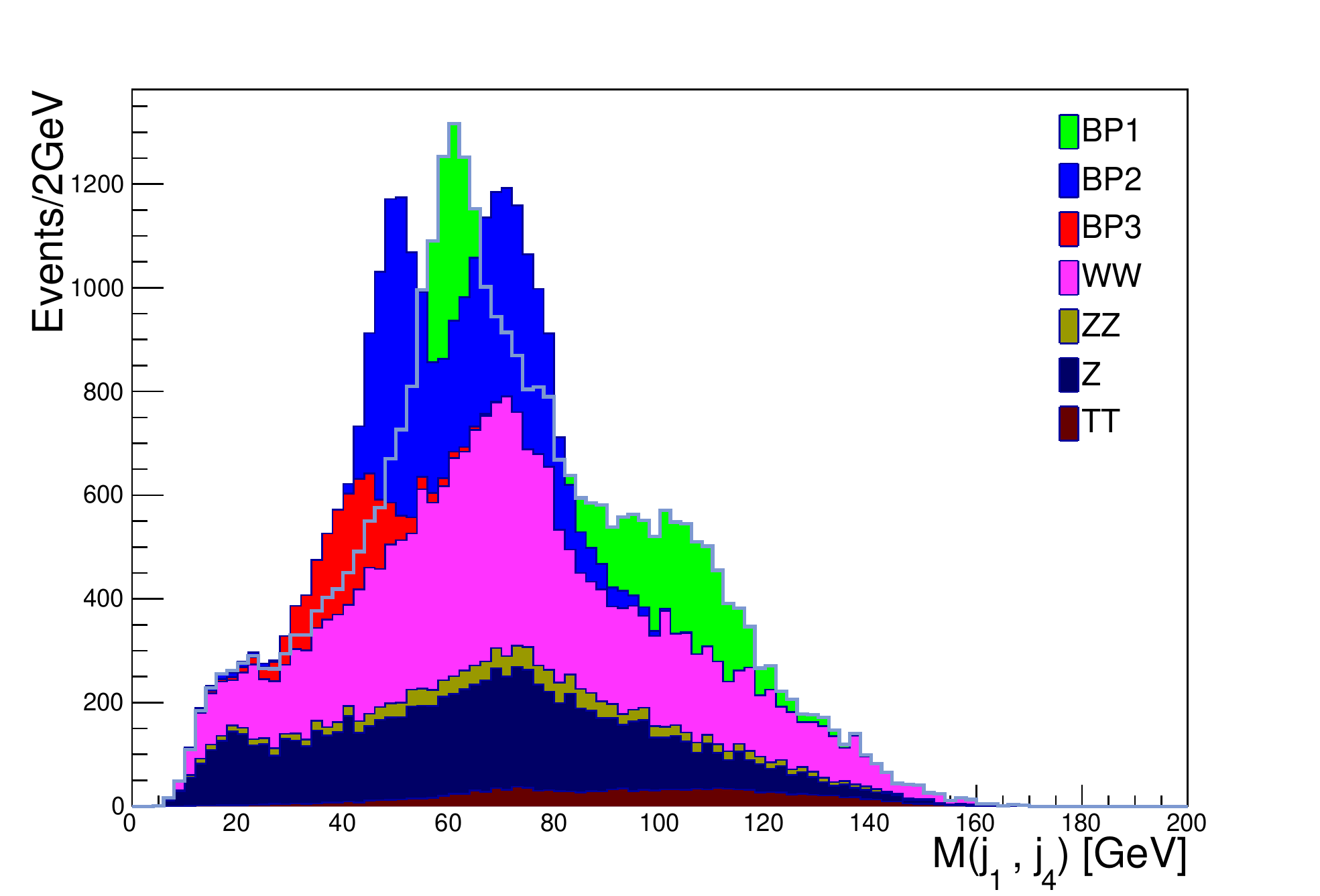}
  \caption{$e^{+}e^{-}\ra jjjjHH$ at $\sqrt{s}=0.5$ TeV}
  \label{mj1j4_jjjjHH}
\end{subfigure}%
\begin{subfigure}{0.5\textwidth}
  \centering
  \includegraphics[width=\textwidth]{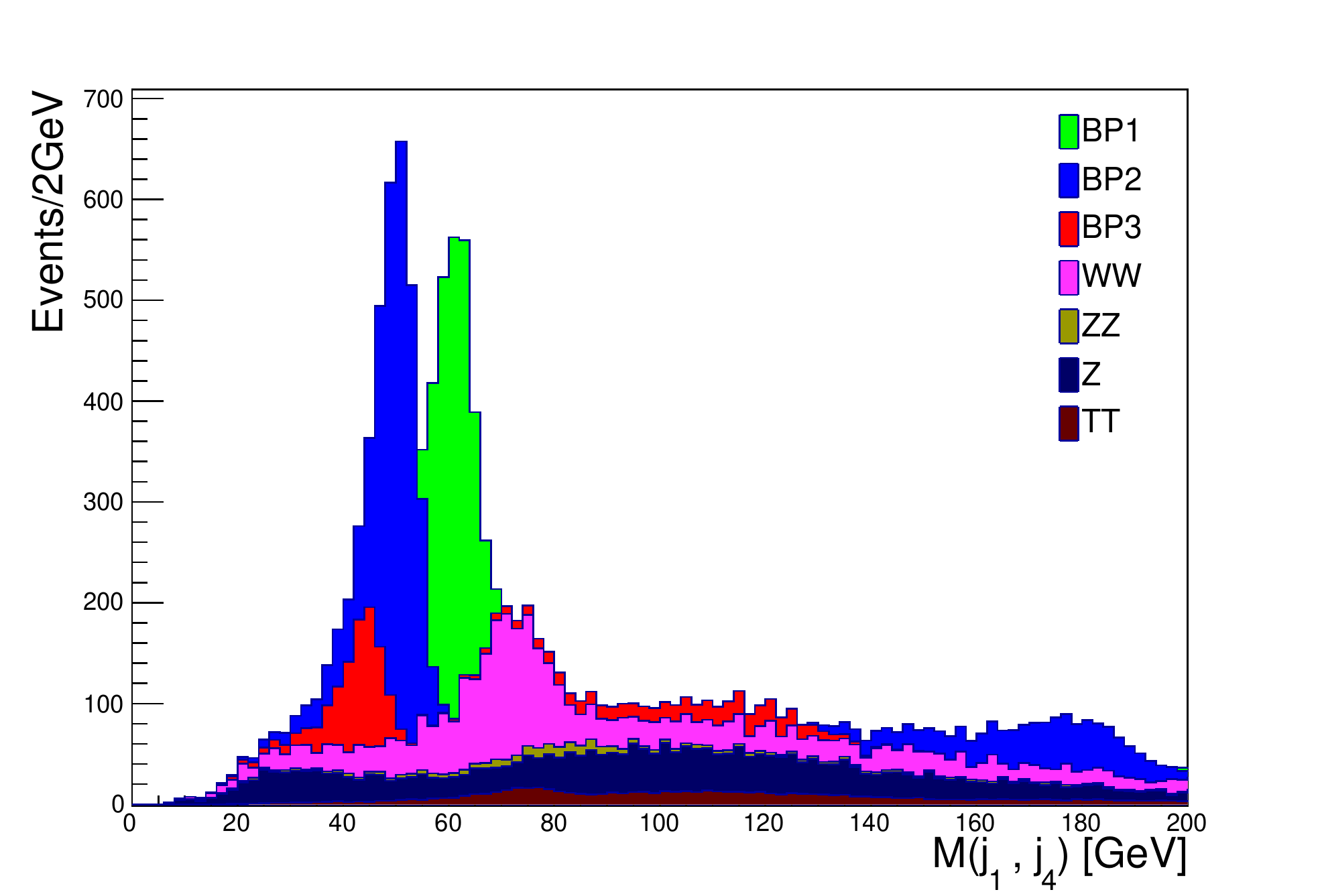}
  \caption{$e^{+}e^{-}\ra jjjjHH$ at $\sqrt{s}=1$ TeV}
  \label{mj1j4_1_jjjjHH}
\end{subfigure}
\caption{Invariant mass of the first and fourth jet in fully hadronic final state at $\sqrt{s}=0.5$ TeV (left) and 1 TeV (right), for $e^+e^-\to H^+H^-$.}
\label{}
\end{figure}

\begin{figure}
\centering
\begin{subfigure}{0.5\textwidth}
  \centering
  \includegraphics[width=\textwidth]{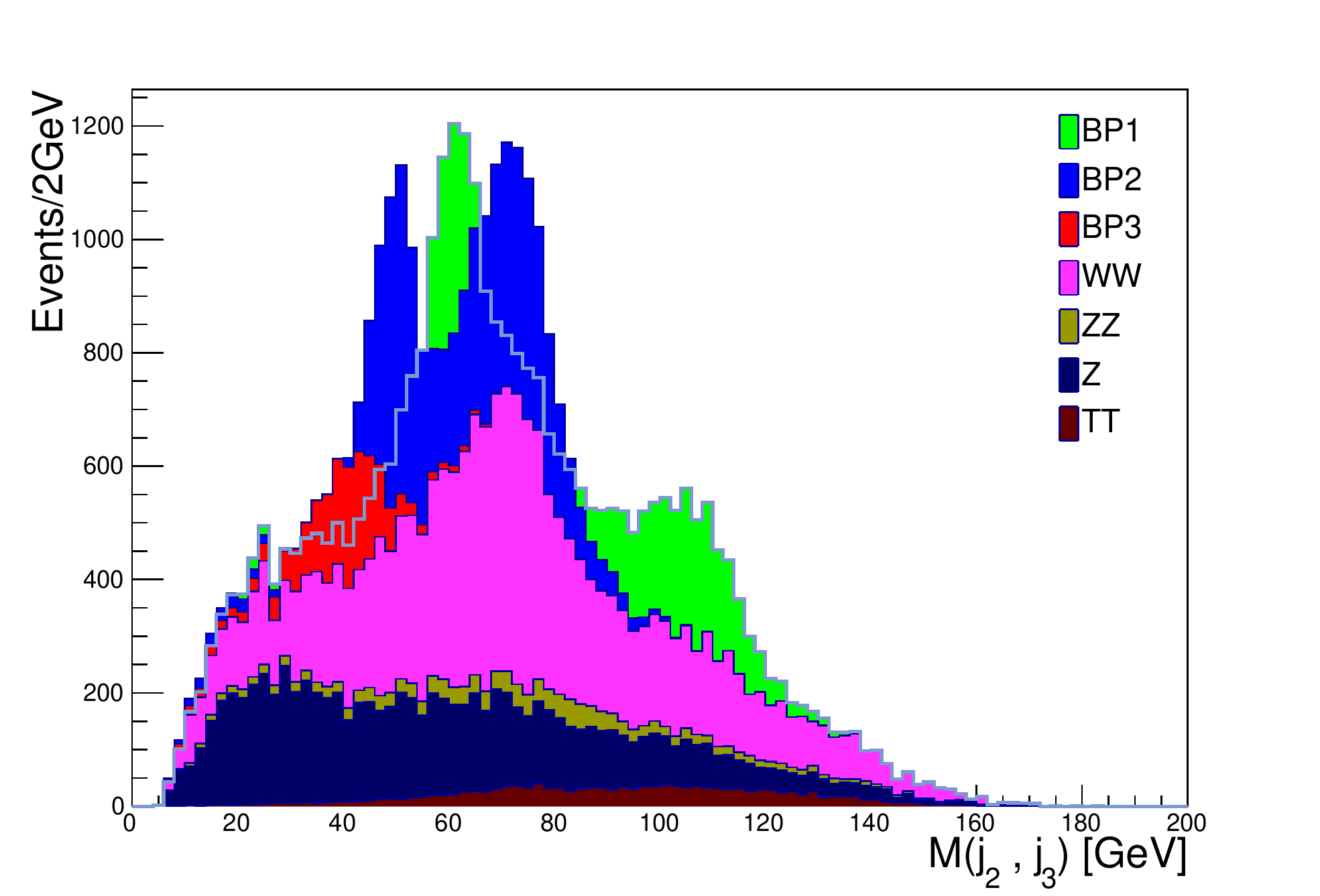}
  \caption{$e^{+}e^{-}\ra jjjjHH$ at $\sqrt{s}=0.5$ TeV}
  \label{mj2j3_jjjjHH}
\end{subfigure}%
\begin{subfigure}{0.5\textwidth}
  \centering
  \includegraphics[width=\textwidth]{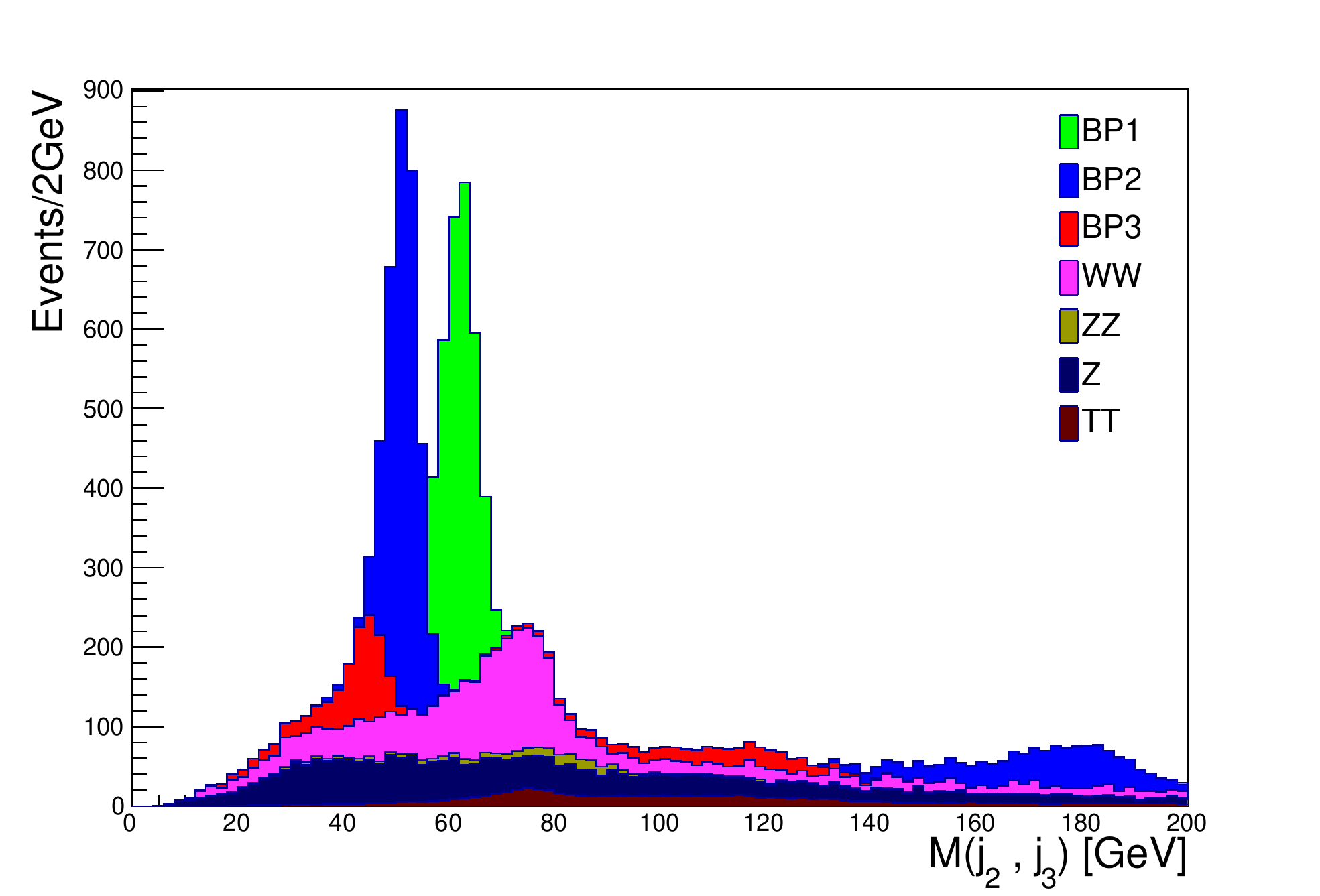}
  \caption{$e^{+}e^{-}\ra jjjjHH$ at $\sqrt{s}=1$ TeV}
  \label{mj2j3_1_jjjjHH}
\end{subfigure}
\caption{Invariant mass of the second and third jet in fully hadronic final state at $\sqrt{s}=0.5$ TeV (left) and 1 TeV (right), for $e^+e^-\to H^+H^-$.}
\label{}
\end{figure}

\Cref{mj1j4_jjjjHH,mj1j4_1_jjjjHH}
show the invariant mass of the first and fourth jet pairs
 ($j_1j_4$) at $\sqrt{s}~=~0.5$ TeV and 1~TeV, respectively. The corresponding
distributions for second and third jet pairs ($j_2j_3$) are shown in
\cref{mj2j3_jjjjHH,mj2j3_1_jjjjHH}, respectively.
As seen in \cref{mj1j4_jjjjHH,mj2j3_jjjjHH} there is a low value
peak which should be related to the right combinations of jets and a
second bump related to wrong matching.
In case of 1 TeV collisions the second bump is much smaller.
These plots are used for signal
extraction using a mass window cut.
\Cref{jjjj_mj1j4_finaleff,jjjj_mj2j3_finaleff} show number of
signal and background events, signal to background ratio and the
signal significance expected for integrated luminosity of
500~fb$^{-1}$, based on a mass window cut on invariant mass of jet
pairs, i.e., $m(j_1j_4)$ and $m(j_2j_3)$ respectively.
If the kinematic fit method was used for selecting proper jet matching,
both invariant mass distributions could be used to get even better
signal to background separation.
However, we have found, that already with the single invariant mass
distribution, very high signal selection significance can be obtained.
The corresponding precision of the signal cross section determination
is of the order of 2--6\%.
Furthermore the position of the peak observed in the invariant mass
distribution can be used to constrain the mass difference
$m_{H^{\pm}}-m_{H}$.
Similar to the semi-leptonic channel, very high statistical precision of the
order of 100~MeV is expected (see discussion in \cref{sec:dm_mass}).

\begin{table}[t]
\centering
\begin{tabular}{||c|c|c|c|c||c|c|c|c||}
\hline\hline
\multicolumn{9}{|c|}{$H^{+}H^{-}$, fully hadronic final state at ${\cal L}=500$~ fb$^{-1}$, cut on $m(j_1j_4)$}\\
\hline
& \multicolumn{4}{c||}{$\sqrt{s}=$0.5 TeV}& \multicolumn{4}{c|}{$\sqrt{s}=$1 TeV}\\
\hline
& S& B& S/B & $S/\sqrt{S+B}$ & S & B & S/B & $S/\sqrt{S+B}$\\
\hline\hline
BP 1 & 4160 & 6742 & 0.61 & 40 & 2967 & 1040 & 2.8 & 47\\
BP 2 & 3751 & 3771 & 0.99 & 43 & 2450 & 473  & 5.2 & 45\\
BP 3 & 1543 & 4040 & 0.38 & 21 & 625  & 668  & 0.9 & 17\\
\hline
\end{tabular}
\caption{Number of events in signal and background processes after all
  selection cuts as in \cref{cuteff3,cuteff4} plus a
  mass window cut on $m(j_1j_4)$ at integrated luminosity of 500
  fb$^{-1}$. S and B stand for the number of signal and background
  events. \label{jjjj_mj1j4_finaleff}}
\end{table}

\begin{table}[t]
\centering
\begin{tabular}{||c|c|c|c|c||c|c|c|c||}
\hline\hline
\multicolumn{9}{|c|}{$H^{+}H^{-}$, fully hadronic final state at ${\cal L}=500$~fb$^{-1}$, cut on $m(j_2j_3)$}\\
\hline
& \multicolumn{4}{c||}{$\sqrt{s}=$0.5 TeV}& \multicolumn{4}{c|}{$\sqrt{s}=$1 TeV}\\
\hline
& S& B& S/B & $S/\sqrt{S+B}$ & S & B & S/B & $S/\sqrt{S+B}$\\
\hline\hline
BP 1 & 3764 & 6210 & 0.6 & 38 & 3137 & 1598 & 2.0 & 45\\
BP 2 & 3445 & 3645 & 0.9 & 41 & 2176 & 855  & 2.5 & 39\\
BP 3 & 1473 & 4374 & 0.3 & 19 & 627  & 1109 & 0.6 & 15\\
\hline
\end{tabular}
\caption{Number of events in signal and background processes after all
  selection cuts as in \cref{cuteff3,cuteff4} plus a
  mass window cut on $m(j_2j_3)$ at integrated luminosity of 500
  fb$^{-1}$. S and B stand for the number of signal and background
  events. \label{jjjj_mj2j3_finaleff}}
\end{table}

\section{$e^{+}e^{-}\rightarrow AH$}
\label{eeHA}

\begin{figure}[t]
\begin{tikzpicture}[node distance=1.5cm,thick, rounded corners=0pt,line cap=round]
\begin{scope}[xshift=0cm]
\coordinate[] (v1);
\coordinate[right=of v1] (v2);
\coordinate[above right=of v2] (v3);
\coordinate[below right=of v2] (o2);
\coordinate[below left=of v1] (i1);
\coordinate[above left=of v1] (i2);
\coordinate[above right=of v3] (v4);
\coordinate[below right=of v3] (o1);
\coordinate[position=15 degrees from v4] (s1);
\coordinate[position=-15 degrees from v4] (s2);
\draw[fermion] (i1)node[left]{$e^-$} -- (v1);
\draw[fermion](v1)--(i2)node[left]{$e^+$};
\draw[boson] (v1) --node[below]{$Z$} (v2);
\draw[scalar] (v3) --node[above left]{$A$} (v2);
\draw[scalar] (v2) --(o2)node[right]{$H$};
\draw[boson] (v3) --node[above left]{$Z$} (v4);
\draw[scalar] (v3) --(o1)node[right]{$H$};
\draw[fermion](s2)node[right]{$\lep$}--(v4);
\draw[fermion](v4)--(s1)node[right]{$\lep$};
\end{scope}
\begin{scope}[xshift=8cm]

\coordinate[] (v1);
\coordinate[right=of v1] (v2);
\coordinate[above right=of v2] (v3);
\coordinate[below right=of v2] (o2);
\coordinate[below left=of v1] (i1);
\coordinate[above left=of v1] (i2);
\coordinate[above right=of v3] (v4);
\coordinate[below right=of v3] (o1);
\coordinate[position=15 degrees from v4] (s1);
\coordinate[position=-15 degrees from v4] (s2);
\draw[fermion] (i1)node[left]{$e^-$} -- (v1);
\draw[fermion](v1)--(i2)node[left]{$e^+$};
\draw[boson] (v1) --node[below]{$Z$} (v2);
\draw[scalar] (v3) --node[above left]{$A$} (v2);
\draw[scalar] (v2) --(o2)node[right]{$H$};
\draw[boson] (v3) --node[above left]{$Z$} (v4);
\draw[scalar] (v3) -- (o1)node[right]{$H$};
\draw[jet](s2)node[right]{$j$}--(v4);
\draw[jet](v4)--(s1)node[right]{$j$};
\end{scope}

\end{tikzpicture}
\caption{The Feynman diagrams for our production and decay processes, $e^+e^-\to H A\to HHZ\to \lep\lep  HH$ and  $e^+e^-\to H A\to HHZ\to  jj HH$.}
\label{fig_feynmanHA}
\end{figure}
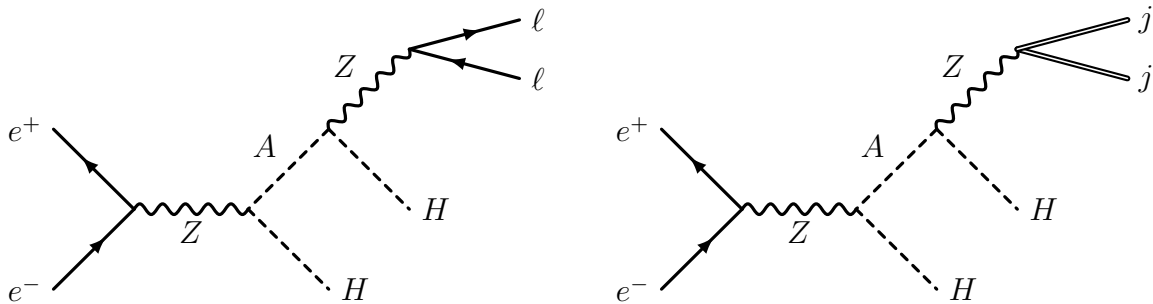

The production of pairs of neutral scalars, presented in this section, is analysed
with the same computational setup as discussed in the previous
section.
In all three benchmark points,
the branching ratio of $A$ to $Z\;H$,  $BR(A\ra ZH)$, is close to 100\%.
Therefore the considered signal events are
$e^{+}e^{-} \ra AH \ra ZHH$.
For the $Z$ boson decay channel, two different final states are taken
into account, i.e. the leptonic final state, where $Z$ decays to a pair
of electrons or muons, and hadronic final state where $Z$ decays to two
jets. Therefore two final states, leptonic and hadronic,
are considered in each mass scenario for center of mass energies
of 0.5 and 1 TeV. \Cref{fig_feynmanHA} shows the signal
production process. We will first present the cross section
calculation for these processes, followed by the detailed description
of event selection and analysis.

\subsection{Signal and Background Cross Sections}

Signal cross section for different scenarios is calculated with CompHep
using Monte Carlo approach. \Cref{sigxsec} shows cross sections
at center of mass energies of 0.5 and 1 TeV. The same set of
background events is used in the analysis  for both final states. However,
events are selected with the final states compatible with the
considered signal process: for leptonic(hadronic) final state analysis, the $ W$ and $Z$  boson
from the background sample are also assumed to decays to leptons (jets).
\begin{table}[t]
\centering
\begin{tabular}{||c|c|c|c||c|c|c||}
\hline
\multirow{2}{*}{Process} &  \multicolumn{6}{c||}{$e^{+}e^{-} \ra AH$}\\ \cline{2-7}
& \multicolumn{3}{c||}{$\sqrt{s}=0.5$ TeV} & \multicolumn{3}{c||}{$\sqrt{s}=1$ TeV}\\
\hline
Benchmark point & BP1 & BP2 & BP3 & BP1 & BP2 & BP3 \\
\hline
Cross section [fb] & 45 & 42.9 & 34.2 & 12.5 & 12.4 &  11.8 \\
\hline
\end{tabular}
\caption{Signal and background cross sections at $\sqrt{s}=0.5$ TeV. \label{sigxsec}}
\end{table}

\subsubsection{Leptonic Final State}

In the leptonic final state analysis signal events contain two
leptons, which are taken to be electrons or muons, and
missing transverse momentum (due to escaping $HH$ pair).
The $WW$ background has to involve $W\ra\lep\nu$ decay
for both $W$ bosons in order to produce two (same flavour) leptons.
In such case, the sum of energies of those leptons is usually higher
than that in the signal events because these leptons stem from high
energy $W$ bosons, while for signal events they come from a single
off-shell $Z$ boson, with much lower Lorentz boost.
For signal events, we also expect to observe the peak in the invariant mass
distribution of the lepton pair, which corresponds to the $m_A-m_H$ mass
difference.

The $ZZ$ background can have one of or both Z bosons decaying to lepton
pairs, i.e., $ZZ \ra \lep\lep jj$ or $ZZ\ra \lep\lep\lep\lep$. The
first type of events can be easily suppressed by a jet veto cut,
while, the second type is reduced by the cut on the number of
leptons.
The $ZZ$ background with one of the bosons decaying to neutrinos
$ZZ \ra \lep\lep \nu \nu$ and the Drell-Yan (single $Z^*/\gamma^*$)
background can be suppressed by rejecting events with  the invariant mass of the
lepton pair is consistent with $Z$ mass.

Summarized in \cref{selcut4} are the  cuts used for
selection of signal events in the leptonic channel.
An event is required to have two leptons with transverse energies
above 1(5) GeV at $\sqrt{s}~=~0.5$(1) TeV. The missing transverse
momentum is required to be in the range $10 ~<~\met ~<~120(250)$ GeV at
$\sqrt{s}~=~0.5$(1) TeV. The lower limit is applied to reject the
Drell-Yan background, while, the upper limit is for suppression of
$WW$ and $ZZ$ events. Finally, the invariant mass of the lepton pair is required to
be outside the mass window of $m_{Z}\pm20$ GeV where $m_{Z}$=90~GeV.
In case of signal events, the invariant mass of the lepton pair is
expected to be peaked at $m_A-m_H$, which is much below $m_{Z}$.
\Cref{cuteff7,cuteff8} show selection efficiencies for
leptonic channel at $\sqrt{s}=0.5$ and 1 TeV, respectively.

\begin{table}[htb]
\centering
\begin{tabular}{|c|c|c|}
\hline
\hline
\multicolumn{3}{|c|}{$HA$ analysis, leptonic final state selection}\\
\hline
Selection cut & $\sqrt{s}=0.5$ TeV & $\sqrt{s}=1$ TeV\\
\hline
2 leptons & $E_{T}>1$ GeV &  $E_{T}>5$ GeV \\
\hline
$\met$ & $10~<~\met~<~120$ GeV & $10~<~\met~<~250$ GeV \\
\hline
$m_{\ell1,\ell2}$ & $|m_{\ell1,\ell2}-m_{Z}|~>~$20 GeV &  $|m_{\ell1,\ell2}-m_{Z}|~>~$20 GeV \\
\hline
\hline
\end{tabular}
\caption{Selection cuts for leptonic final state analysis at two center of mass energies of 0.5 and 1 TeV.\label{selcut4}}
\end{table}

\begin{table}[h]
\centering
\begin{tabular}{|c|c|c|c|c|c|c|c|}
\hline
\hline
\multicolumn{8}{|c|}{$HA$ analysis, leptonic final state selection}\\
\hline
Cut eff. & BP1 & BP2 & BP3 & WW & ZZ & Z & TT \\
\hline
Two Leptons & 0.99 & 1 & 0.92 & 0.76 & 0.083 & 0.8 & 0.4\\
\hline
$\met$ & 1 & 1 & 0.24 & 0.91 & 0 & 3.9e-05 & 0.89\\
\hline
Z suppression & 1 & 1 & 1 & 0.97 & -  & 0.48 & 0.73\\
\hline
Total eff. & 0.99 & 1 & 0.22 & 0.67 & 0 & 1.5e-05 & 0.26\\
\hline
\hline
\end{tabular}
\caption{Cut efficiencies for leptonic final state analysis at the center of mass energy of 0.5 TeV.\label{cuteff7}}
\end{table}
\begin{table}[h]
\centering
\begin{tabular}{|c|c|c|c|c|c|c|c|}
\hline
\hline
\multicolumn{8}{|c|}{$HA$ analysis, leptonic final state selection}\\
\hline
Cut eff. & BP1 & BP2 & BP3 & WW & ZZ & Z & TT \\
\hline
Two Leptons & 0.98 & 0.98 & 0.65 & 0.5 & 0.19 & 0.66 & 0.52\\
\hline
$\met$ & 1 & 1 & 1 & 0.92 & 2.1e-05 & 0.0001 & 0.96\\
\hline
Z suppression & 1 & 1 & 1 & 0.98 & 0.5 & 0.63 & 0.86\\
\hline
Total eff. & 0.98 & 0.98 & 0.65 & 0.45 & 2e-06 & 4.2e-05 & 0.42\\
\hline
\hline
\end{tabular}
\caption{Cut efficiencies for leptonic final state analysis at the center of mass energy of 1 TeV.\label{cuteff8}}
\end{table}

\begin{figure}
\centering
\begin{subfigure}{0.5\textwidth}
  \centering
  \includegraphics[width=\textwidth]{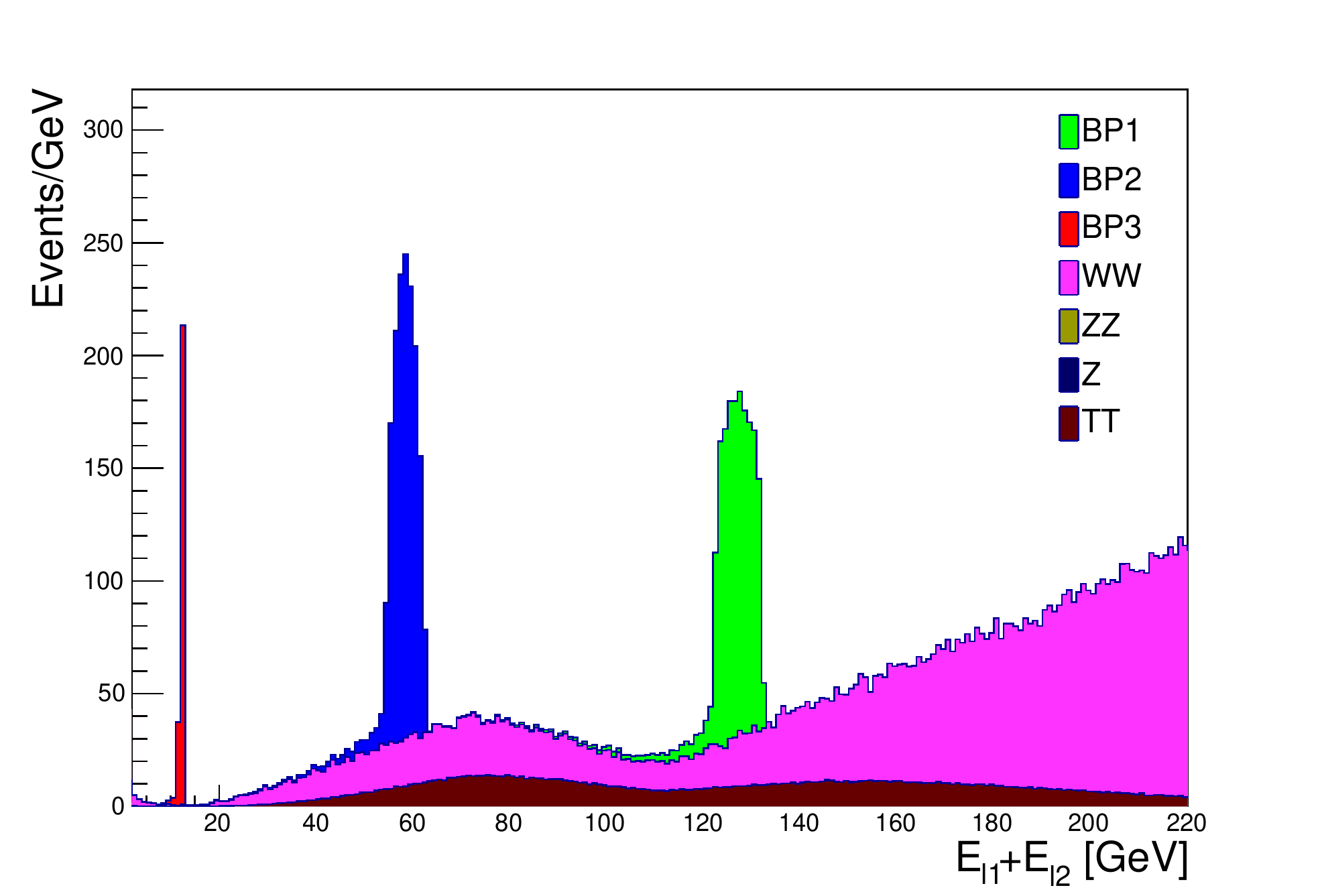}
  \caption{$e^{+}e^{-}\ra \ell\ell HH$ at $\sqrt{s}=0.5$ TeV}
  \label{el1l2_AHllHH}
\end{subfigure}%
\begin{subfigure}{0.5\textwidth}
  \centering
  \includegraphics[width=\textwidth]{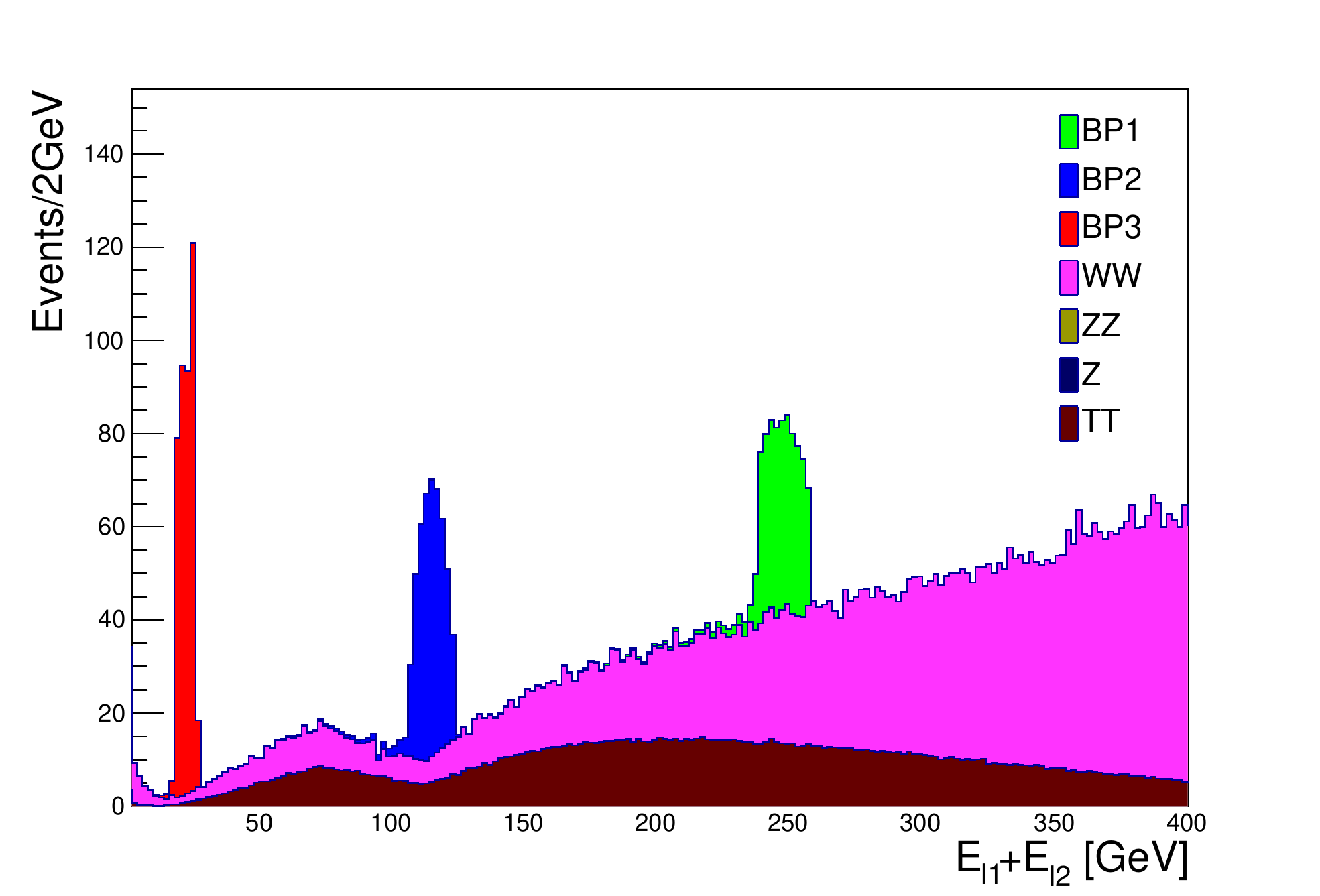}
  \caption{$e^{+}e^{-}\ra \ell\ell HH$ at $\sqrt{s}=1$ TeV}
  \label{el1l2_1_AHllHH}
\end{subfigure}
\caption{Sum of energies of two leptons in leptonic final state at $\sqrt{s}=0.5$ TeV (left) and 1 TeV (right), for $e^+e^-\to H A$
.}
\label{}
\end{figure}

\begin{figure}
\centering
\begin{subfigure}{0.5\textwidth}
  \centering
  \includegraphics[width=\textwidth]{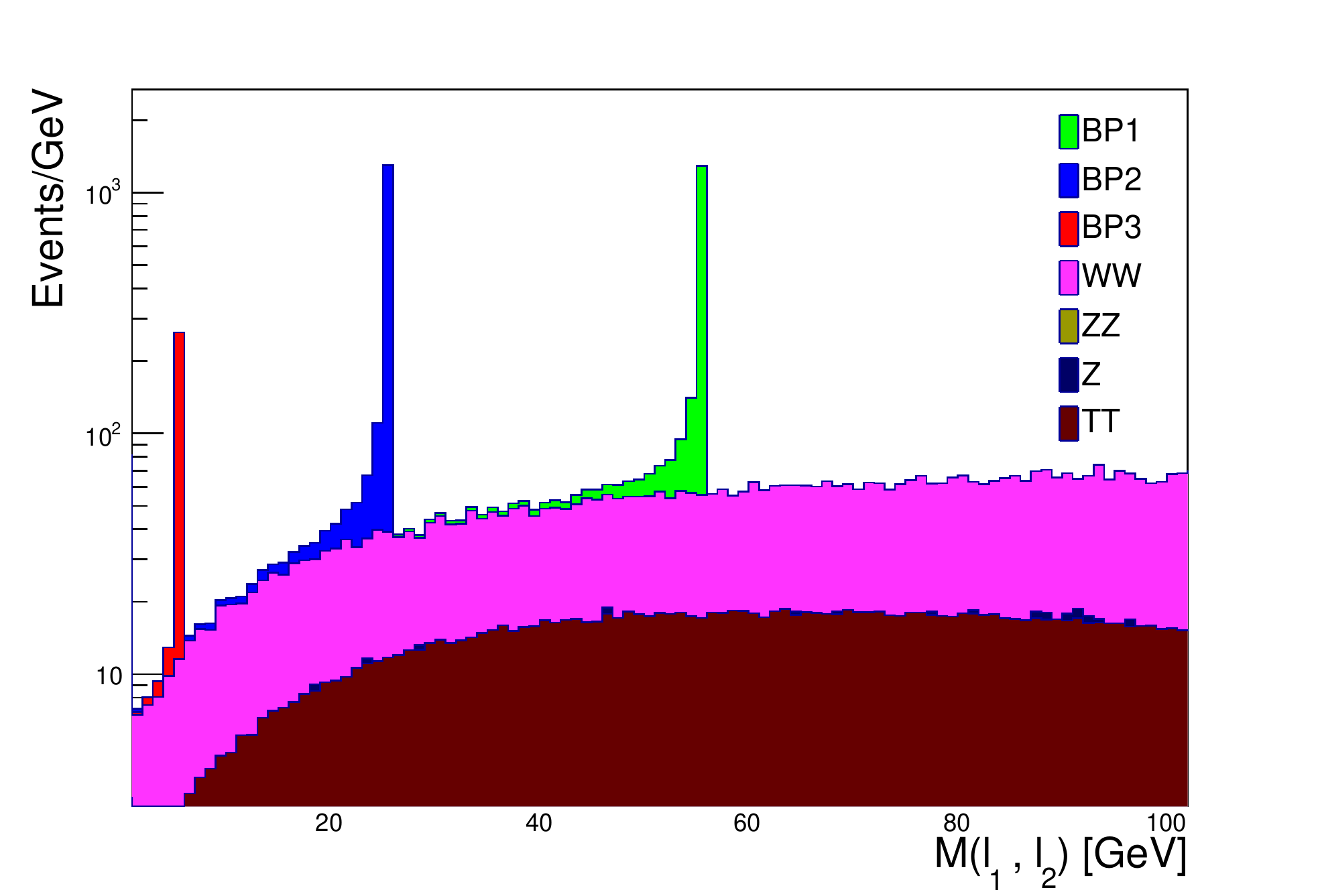}
  \caption{$e^{+}e^{-}\ra \ell\ell HH$ at $\sqrt{s}=0.5$ TeV}
  \label{ml1l2_AHllHH}
\end{subfigure}%
\begin{subfigure}{0.5\textwidth}
  \centering
  \includegraphics[width=\textwidth]{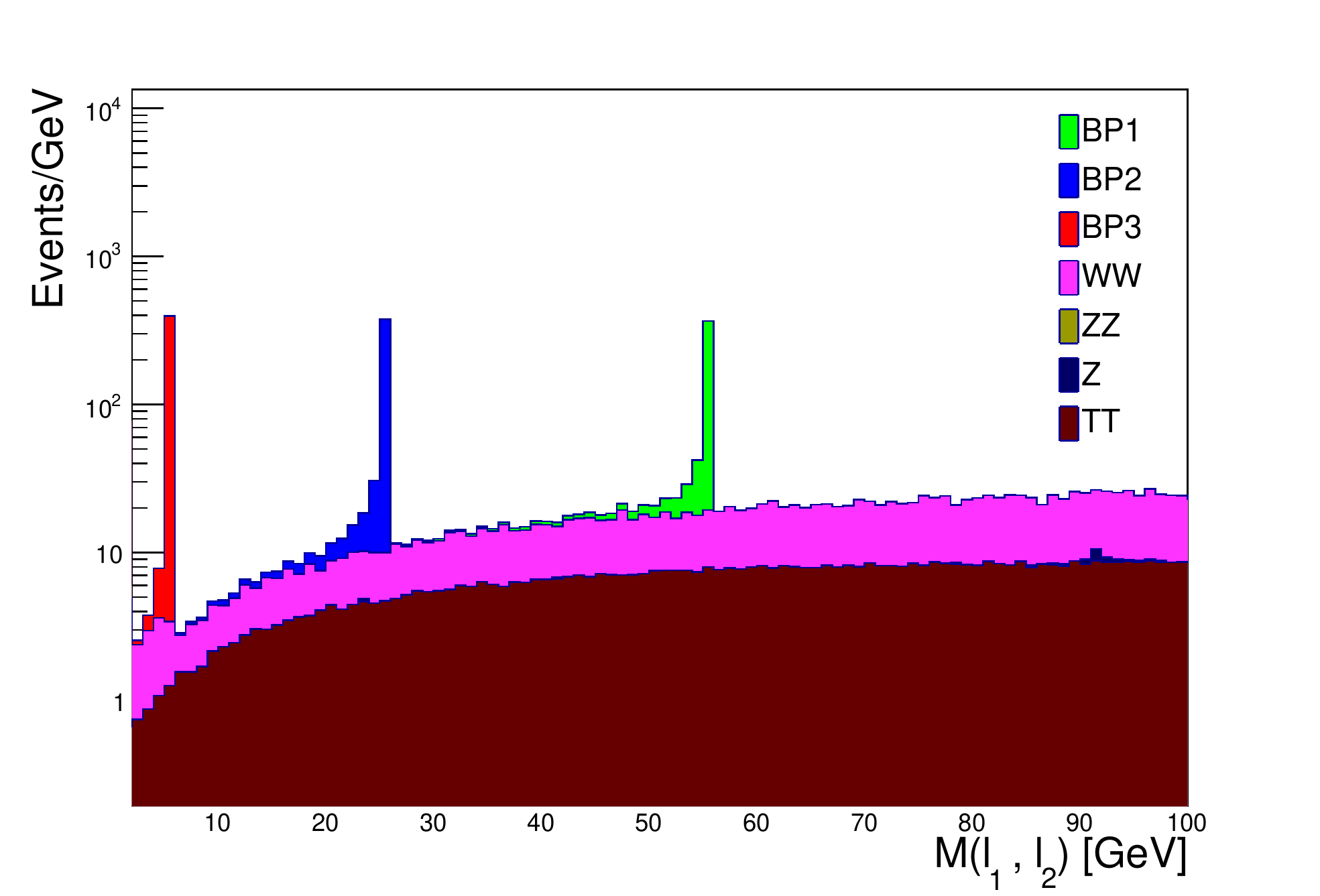}
  \caption{$e^{+}e^{-}\ra \ell\ell HH$ at $\sqrt{s}=1$ TeV}
  \label{ml1l2_1_AHllHH}
\end{subfigure}
\caption{Invariant mass of the lepton pair in leptonic final state at $\sqrt{s}=0.5$ TeV (left) and 1 TeV (right), for $e^+e^-\to H A$.}
\label{}
\end{figure}

The final signal event selection may be based on the sum of energies of the
lepton pair (\cref{el1l2_AHllHH,el1l2_1_AHllHH}) or on the
invariant mass distribution (\cref{ml1l2_AHllHH,ml1l2_1_AHllHH}),
since both show clear separation between signal and background processes.
As leptons are reconstructed with a high efficiency and very good momentum
resolution, a very sharp edge should be observed in the invariant mass
distributions, which can be used to reconstruct the value of $m_A-m_H$
with negligible statistical uncertainty (see discussion in
\cref{sec:dm_mass}).
Applying a mass window cut on invariant mass distributions of
\cref{ml1l2_AHllHH,ml1l2_1_AHllHH}, the numbers of
signal and background events are counted to estimate the signal
significance as shown in \cref{llHH}.
The corresponding precision of the signal cross section determination
is 3--7\%.

\begin{table}[t]
\centering
\begin{tabular}{||c|c|c|c|c||c|c|c|c||}
\hline\hline
\multicolumn{9}{|c|}{$HA$, leptonic final state at ${\cal L}=500$~fb$^{-1}$}\\
\hline
& \multicolumn{4}{c||}{$\sqrt{s}=$0.5 TeV}& \multicolumn{4}{c|}{$\sqrt{s}=$1 TeV}\\
\hline
& S& B& S/B & $S/\sqrt{S+B}$ & S & B & S/B & $S/\sqrt{S+B}$\\
\hline
\hline
BP 1 & 1382 & 279 & 4.9 & 34 & 386 & 55 & 7& 18 \\
\hline
BP 2 & 1378 & 186 & 7.4 & 35 & 397 & 30 & 13 & 19\\
\hline
BP 3 & 256 & 81 & 3.1 & 14   & 257 & 51 & 5& 15\\
\hline
\end{tabular}
\caption{Number of events in signal and background processes after all selection cuts at integrated luminosity of 500~fb$^{-1}$. \label{llHH}}
\end{table}


\subsubsection{Hadronic Final State}

In the hadronic final state, signal events contain two jets from the off-shell
$Z$ decay and missing transverse momentum from the escaping $HH$ pair.
The analysis follows the same strategy, as described above for the leptonic
final state.

An event is required to have two jets reconstructed with transverse
energies above 5~GeV. The missing transverse momentum is
required to be in the range $10~<~\met~<~120(250)$~GeV, for
$\sqrt{s}~=~0.5(1)$ TeV, following the same reasoning as described
above for the leptonic final state.
However, the precision of the invariant mass reconstruction is much
poorer for two jets than it is for two leptons.
Therefore, instead of a cut on the invariant mass of the jet pair,
we require that the sum of jet energies
is less than 150(300) GeV at $\sqrt{s}~=~0.5(1)$ TeV.
This constrain replaces the cut on the invariant mass of the jet pair and
real $Z$ boson suppression cut.
\Cref{selcut5} lists a summary of selection cuts for the hadronic final
state while  \cref{cuteff9,cuteff10} show selection efficiencies at $\sqrt{s}=0.5$ and 1 TeV, respectively. High signal selection efficiency is obtained, except for the BP3 scenario which can hardly be observed in the hadronic channel at $\sqrt{s}=0.5$ TeV.

\begin{figure}[tb]
\centering
\begin{subfigure}{0.5\textwidth}
  \centering
  \includegraphics[width=\textwidth]{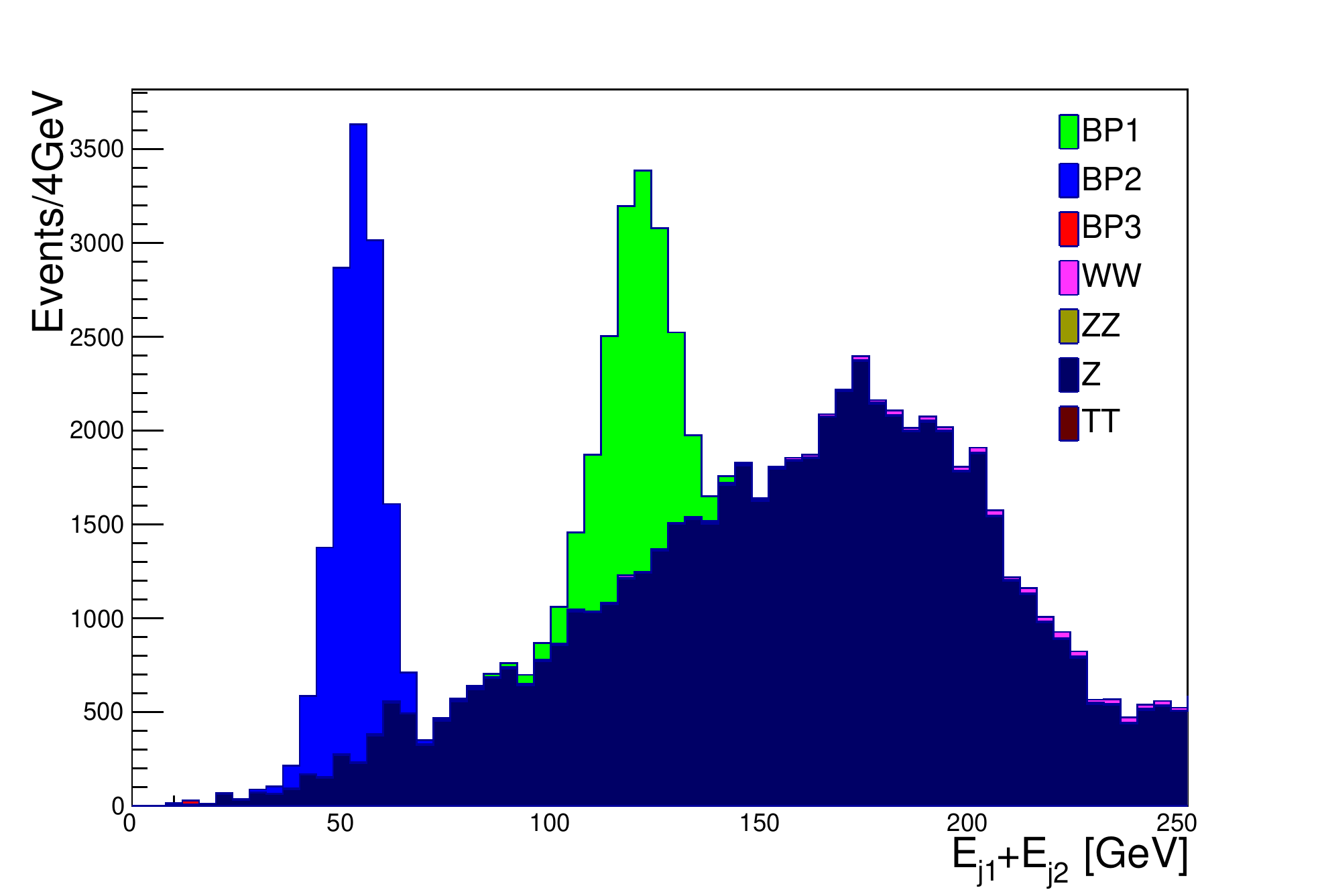}
  \caption{$e^{+}e^{-}\ra jjHH$ at $\sqrt{s}=0.5$ TeV}
  \label{ej1j2_AHjjHH}
\end{subfigure}%
\begin{subfigure}{0.5\textwidth}
  \centering
  \includegraphics[width=\textwidth]{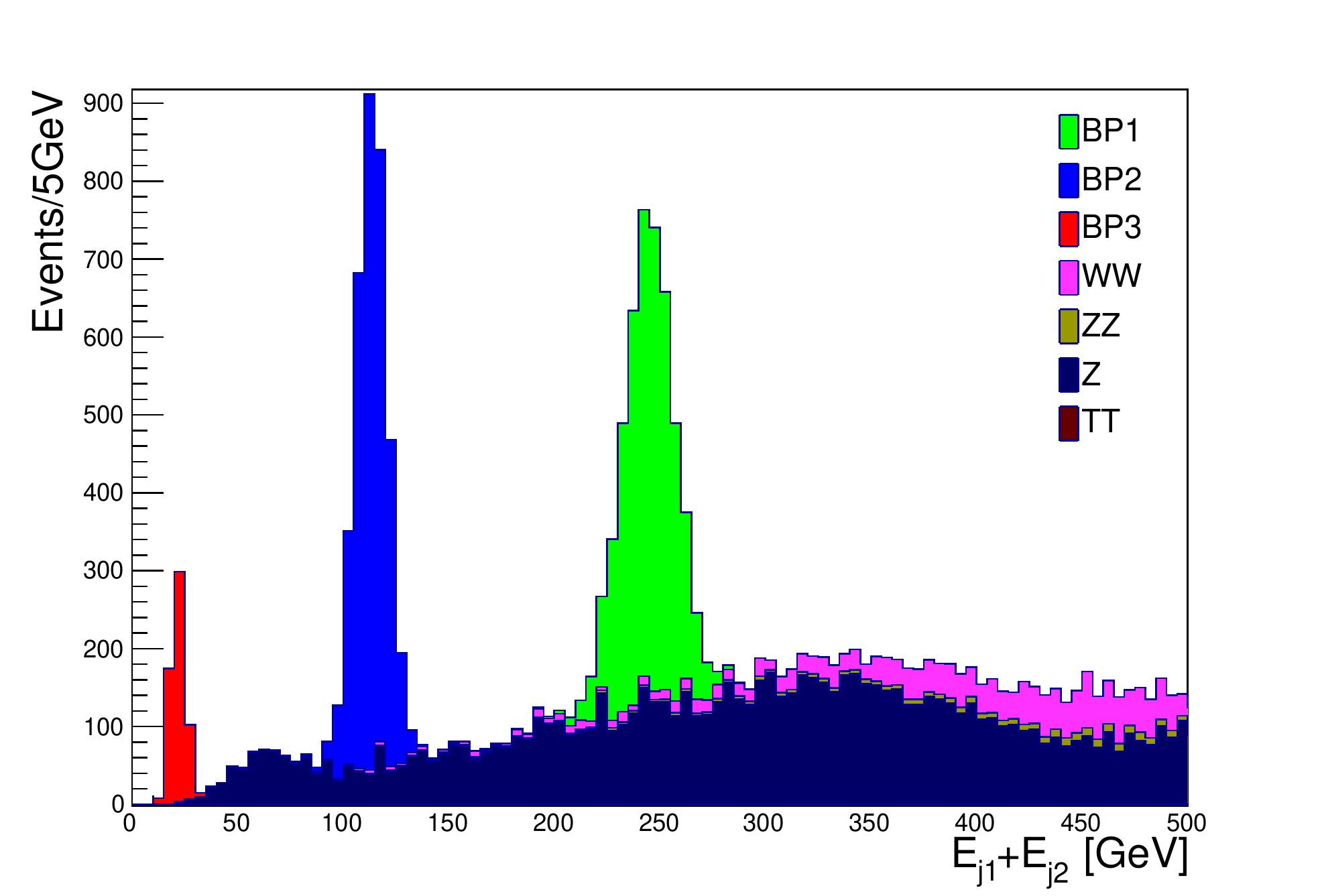}
  \caption{$e^{+}e^{-}\ra jjHH$ at $\sqrt{s}=1$ TeV}
  \label{ej1j2_1_AHjjHH}
\end{subfigure}
\caption{Sum of two jets energies in hadronic final state at $\sqrt{s}=0.5$ TeV (left) and 1 TeV (right), for $e^+e^-\to H A$.}
\label{}
\end{figure}

\begin{figure}[tb]
\centering
\begin{subfigure}{0.5\textwidth}
  \centering
  \includegraphics[width=\textwidth]{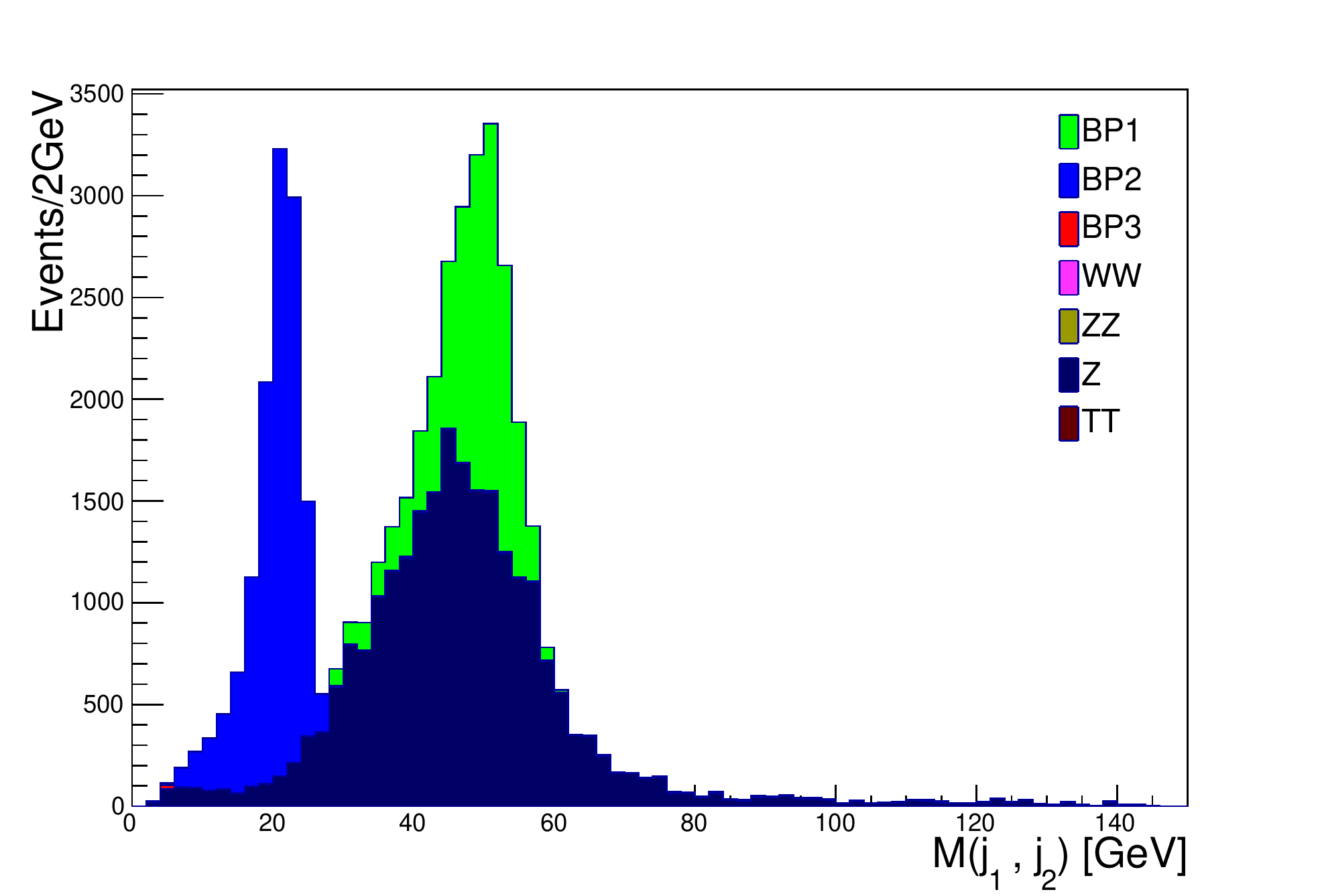}
  \caption{$e^{+}e^{-}\ra jjHH$ at $\sqrt{s}=0.5$ TeV}
  \label{mj1j2_AHjjHH}
\end{subfigure}%
\begin{subfigure}{0.5\textwidth}
  \centering
  \includegraphics[width=\textwidth]{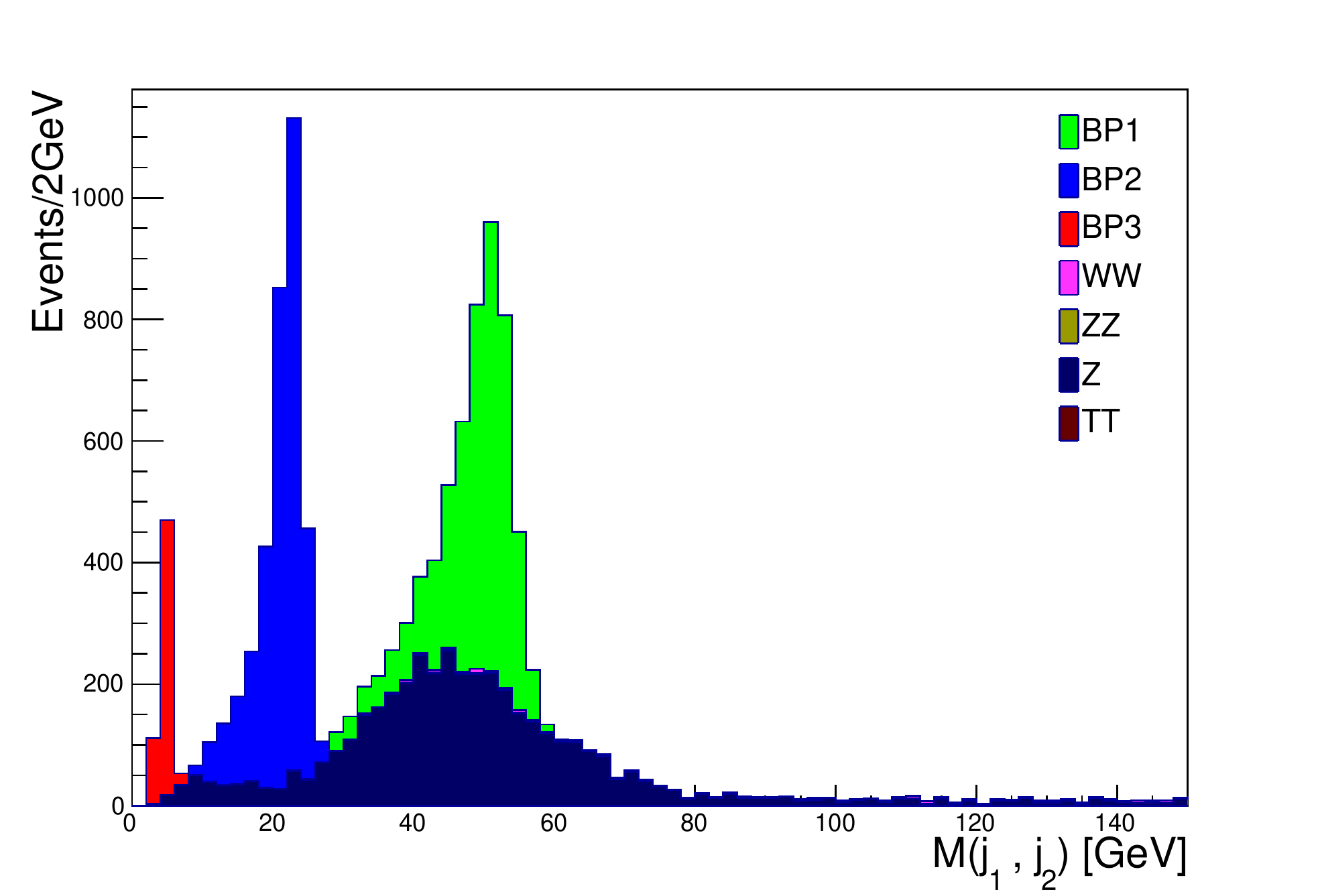}
  \caption{$e^{+}e^{-}\ra jjHH$ at $\sqrt{s}=1$ TeV}
  \label{mj1j2_1_AHjjHH}
\end{subfigure}
\caption{Invariant mass of the two jets in hadronic final state at $\sqrt{s}=0.5$ TeV (left) and 1 TeV (right), for $e^+e^-\to H A$.}
\label{}
\end{figure}

\Cref{ej1j2_AHjjHH,ej1j2_1_AHjjHH} show the
distributions of the sum of jet energies in signal and background
events. These plots justify the cut on the energy sum described
above. The invariant mass distributions of
the jet pairs are shown in \cref{mj1j2_AHjjHH,mj1j2_1_AHjjHH}.
These distributions are used for the final event selection with a
mass window cut.  \Cref{jjHH} summarizes the results obtained by
counting the number of signal and background events after final
selection and mass window cuts.

\begin{table}[htb]
\centering
\begin{tabular}{|c|c|c|}
\hline
\hline
\multicolumn{3}{|c|}{$HA$ analysis, hadronic final state selection}\\
\hline
Selection cut & $\sqrt{s}=0.5$ TeV & $\sqrt{s}=1$ TeV\\
\hline
2 jets & $E_{T}>5$ GeV &  $E_{T}>5$ GeV \\
\hline
$\met$ & $10~<~\met~<~120$ GeV & $10~<~\met~<~250$ GeV \\
\hline
$E(j_1)+E(j_2)$ & $E(j_1)+E(j_2)~<~150$ GeV & $E(j_1)+E(j_2)~<~300$ GeV \\
\hline
\hline
\end{tabular}
\caption{Selection cuts for hadronic final state analysis at two center of mass energies of 0.5 and 1 TeV.\label{selcut5}}
\end{table}

\begin{table}[htb]
\centering
\begin{tabular}{|c|c|c|c|c|c|c|c|}
\hline
\hline
\multicolumn{8}{|c|}{$HA$ analysis, hadronic final state selection}\\
\hline
Cut eff. & BP1 & BP2 & BP3 & WW & ZZ & Z & TT \\
\hline
Two Jets & 0.67 & 0.78 & 0.0027 & 0.033 & 0.036 & 0.36 & 2e-06\\
\hline
$\met$ & 1 & 1 & 0.91 & 0.018 & 0.066 & 0.061 & 0\\
\hline
$E(j_1)+E(j_2)$ &1 & 1 & 1 & 0.16 & 0.19 & 0.19 & 0\\
\hline
Total eff. & 0.67 & 0.78 & 0.0025 & 9.5e-05 & 0.00045 & 0.0041 & 0\\
\hline
\hline
\end{tabular}
\caption{Cut efficiencies for hadronic final state analysis at center of mass energy of 0.5 TeV.\label{cuteff9}}
\end{table}

\begin{table}[htb]
\centering
\begin{tabular}{|c|c|c|c|c|c|c|c|}
\hline
\hline
\multicolumn{8}{|c|}{$HA$ analysis, hadronic final state selection}\\
\hline
Cut eff. & BP1 & BP2 & BP3 & WW & ZZ & Z & TT \\
\hline
Two Jets & 0.87 & 0.76 & 0.14 & 0.22 & 0.14 & 0.24 & 2.5e-05\\
\hline
$\met$ & 1 & 1 & 1 & 0.028 & 0.089 & 0.073 & 0.2\\
\hline
$E(j_1)+E(j_2)$ & 1 & 1 & 1 & 0.078 & 0.14 & 0.17 & 0\\
\hline
Total eff. & 0.87 & 0.76 & 0.14 & 0.00049 & 0.0018 & 0.003 & 0\\
\hline
\hline
\end{tabular}
\caption{Cut efficiencies for hadronic final state analysis at center of mass energy of 1 TeV.\label{cuteff10}}
\end{table}

\begin{table}[htb]
\centering
\begin{tabular}{||c|c|c|c|c||c|c|c|c||}
\hline\hline
\multicolumn{9}{|c|}{$HA$, hadronic final state at ${\cal L}=500$~fb$^{-1}$}\\
\hline
& \multicolumn{4}{c||}{$\sqrt{s}=$0.5 TeV}& \multicolumn{4}{c|}{$\sqrt{s}=$1 TeV}\\
\hline
& S& B& S/B & $S/\sqrt{S+B}$ & S & B & S/B & $S/\sqrt{S+B}$\\
\hline\hline
BP 1 & 3972 & 3926 & 1.0 & 45 & 2693 & 1074 & 2.5 & 44\\
\hline
BP 2 & 7011 & 709 & 9.9 & 80  & 2880 & 199 & 14 & 52\\
\hline
BP 3 & 30 & 195 & 0.15 & 2    & 399 &  8 & 49 & 20\\
\hline
\end{tabular}
\caption{Number of events in signal and background processes after all selection cuts at integrated luminosity of 500~fb$^{-1}$. S and B stand for the number of signal and background events. \label{jjHH}}
\end{table}

\section{Dark Matter Mass Measurement}
\label{sec:dm_mass}

As shown above, the energy and invariant mass distributions for
signal events show clear peaks, which result from the kinematic
constraints of the considered scenario and can be related to the scalar masses.
In this section, we propose a procedure for determination of charged and
neutral scalar masses and estimate the statistical precision, which
can be reached at $e^+ e^-$ collider with 500~fb$^{-1}$.
Feasibility of mass and spin measurement of IDM scalars, in particular
of the DM candidate, were already studied for the  $e^+ e^-$ collider,
see e.g. \cite{Ginzburg:2014ora}.
However, the approach presented in \cite{Ginzburg:2014ora} focuses on
the single particle distributions,

  which we consider challenging from the experimental point of view.
  In particular, lepton energy distributions have to be
  measured down to very low energies, of the order of few GeV.
  It requires efficient identification of low energy leptons and very
  good background understanding.

We propose an approach, which makes use of the
reconstructed peaks in the energy and invariant mass distributions.
The approach is based on the observation that the off-shell $W^*$ and
$Z^*$ bosons are most likely produced with their virtualities close to the
maximum allowed values given by the mass differences $m_{H^{\pm}}-m_{H}$ and
$m_{A}-m_{H}$, respectively.

  First, we consider the distribution of the sum of energies of
  the two jets, in   the semi-leptonic final state  of charged scalar
  production, $\sum_{i=1}^{2}E(j_i)$.
  In the $W^*$ boson rest frame, the sum of jet energies is equal to the
  $W^*$ boson mass and its most probable value is given by
  $m_{H^{\pm}}-m_{H}$.
  When produced with close to maximal virtuality, the $W^*$ boson is almost
  at rest in the reference frame of decaying $H^{\pm}$.
  Therefore the Lorentz boost applied to jet energies can be
  approximated by the $H^\pm$ boost.
  It can be shown that in this approximation
\beq
\sum\limits_{i=1}^{2}E(j_i)=E_{beam}\left(1-\frac{m_{H}}{m_{H^{\pm}}}\right)
\; .
\label{eqk1}
\eeq
Defining $R=\frac{m_{H}}{m_{H^{\pm}}}$, one can solve \cref{eqk1}
to obtain
\beq
R=1-\frac{\sum\limits_{i=1}^{2}E(j_i)}{E_{beam}} \; .
\label{eq_R}
\eeq
To reconstruct the most probable value of the jet energy sum, which
should be use in \cref{eq_R}, the distribution was fitted with a
Gaussian function assuming the background probability density
function (p.d.f) is well known from simulation.
\Cref{eq_R} can be then used to calculate the values of $m_H/m_H^{\pm}$
for each channel.
In case of the four jet final state (fully hadronic final state for
charged scalar pair production), the four jet energy sum can be
divided by two to get the proper estimate of the jet pair energy.
An average value of $R$ can be then calculated, based on all
considered channels.

  In the next step, we consider the invariant mass distributions
  for the two jets of  the semi-leptonic final state in charged scalar
  production, $m(j_1,j_2)$.
  As already mentioned above, this distribution is expected to peak at
  the most probable $W^*$ boson virtuality, which is
\beq
m(j_1,j_2) = m_{H^{\pm}}-m_{H} \; .
\eeq
  As before, we apply a Gaussian fit on signal plus background
  distributions to obtain the signal peak position. This procedure
  can be also used for two jet or two lepton invariant mass
  distribution for the neutral scalar pair production events, providing
  the value of $m_A-m_H$.
  Using the average value of $m_H/m_{H^{\pm}}$ (denoted as $R$)
  obtained from the first step and $m_{H^{\pm}}-m_H$ average value
  from the second step (based on charged scalar production analysis
  results), we can extract $H$ and $H^{\pm}$ masses for each
  considered scenario.

  Finally, we consider the invariant mass distributions
  for the two leptons of the leptonic final state and for the two jets
  of  the hadronic final state, in neutral scalar production process.
  Both distributions are expected to peak at the most probable $Z^*$
  virtuality, which is
\beq
m(l_1,l_2) = m(j_1,j_2) = m_{A}-m_{H} \; .
\eeq
  Due to the very good track momentum resolution, much more precise
  invariant mass determination is expected in the leptonic channel.
  Based on the expected performance of the tracking system for ILC and
  CLIC detectors \cite{Behnke:2013lya,Aicheler:2012bya},
  we estimate the invariant
  mass resolution in the leptonic channel to be of the order of
  0.1-0.2~GeV, resulting in the statistical precision of the mass
  difference determination of the order of 10~MeV.
  As the $H$ boson mass can be known from the charged scalar production
  analysis, value of $m_A-m_H$  extracted from the invariant mass
  distributions for the neutral scalar  pair production can be used to
  calculate the value of $m_A$.

The procedure described above allows for evaluation of all inert scalar
masses, i.e. for the full reconstruction of the IDM spectrum.
However, there are additional constraints which can be used to test
the obtained results, based on the single lepton or single jet energy
distribution or the total energy distribution in the neutral scalar pair
production events.
Energy distribution for single lepton or single jet from the
  semi-leptonic or fully hadronic decay channel, in the charged scalar
  pair production, is expected to be flat. However, the maximum allowed
  energy can be related to scalar masses
\beq
E_{l/j}^{max}=\frac{E_{beam}}{2}(1-R)\left(1+\sqrt{1-\frac{m_{H^{\pm}}^{2}}{E_{beam}^{2}}}\right).
\label{eq_Ej}
\eeq
  \Cref{eq_Ej} can be used for independent evaluation of
  $m_{H^{\pm}}$ when $R$ value is known. However, determination of the
  threshold position in the presence of the significant background is
  an experimental challenge. It requires very good background and
  detector modeling and We do not consider this measurement in the
  presented analysis.

 The $m_{A}$ can also be extracted from the neutral scalar pair
  production events, from  energies of leptons or jets coming from
  $Z^*$ decay.  In the laboratory frame the relation between the sum
  of energies of leptons or jets and neutral scalar masses can be
  written as:
  \beq
E_{\lep\lep \; / \; jj} = \frac{m_{A}-m_{H}}{m_{A}}\sqrt{m_{A}^2+\frac{(s-(m_{A}+m_{H})^2)(s-(m_{A}-m_{H})^2)}{4s}}
\label{eq_Ez}
\eeq
However, the value of $m_{A}$ extracted from this equation turns out
to be much more sensitive to the value of $m_H$ than for the method
based on the invariant mass measurement.

Following the steps described above, masses of all charged and neutral
scalars can be obtained with a statistical precision of the order
100~MeV, as shown in \cref{masses}.
The systematic shifts observed between the assumed (theo.)  scalar masses and
the values resulting from the calculations  are due to the simplified approach
used, but can be corrected based on the simulation results.

\begin{table}[p]
  \centering
  {\small
    \include{IDM_table21}
    }
\caption{Positions of the reconstructed peaks in the
energy and invariant mass distributions for the charged scalar (a)
and neutral scalar (b) pair production, and the
reconstructed inert scalar masses (c).
Different decay channels are considered for center of mass energies of
0.5 and 1 TeV, as indicated in the table.
Results on the scalar mass differences, $m_{H^{\pm}}-m_H$ and
$m_{A}-m_H$, and mass ratio $R=m_H/m_{H^{\pm}}$ are first averaged
over different final states and then used for scalar mass reconstruction as
described in the text.
Errors indicated correspond to the statistical uncertainties only.
\label{masses}}
\end{table}

\section{Conclusions}
\label{Conclusion}

The Inert Doublet Model was studied as the underlying theoretical
framework for light charged and neutral dark scalar production at $e^{+}e^{-}$
colliders. For the charged scalar production, a pair production
through $e^{+}e^{-} \ra H^{+}H^{-}$ was taken as the signal, while for
neutral scalars production, $e^{+}e^{-} \ra AH$ was considered.
Three  benchmark scenarios with scalar masses below 200~GeV,
obtained recently in \cite{Ilnicka:2015jba},   were tested and detailed
analyses were designed for each considered production channel and final state.
Results of the analyses show that, for the considered IDM benchmark
scenarios, production of dark scalars should be observable already at
the early stages of   $e^+e^-$ colliders running at center of mass energy
of either 0.5 or 1 TeV.
With 500~fb$^{-1}$ of data, the signal cross section can be measured
with precision between 2\% and 12\%, depending on the considered scenario and decay channel.
Using the reconstructed invariant mass and energy distributions of the
visible decay products, the masses of dark matter particles can
be extracted with a negligible statistical precision, of the order of 100~MeV.
Therefore, we expect that precision of the IDM dark scalar mass
measurement at the future $e^{+}e^{-}$ collider will be dominated by
systematic effect.


\section*{Acknowledgements}

The authors would like to thank Ilya Ginzburg for
useful discussions and Tania Robens for useful comments and
especially for critical reading of the manuscript.
M.H would like to thank Dr. Mogharrab for the
operation and maintenance of the computing cluster at Shiraz
university. This research was supported by the  National Science
Centre (Poland) within an OPUS research project 2012/05/B/ST2/03306
(2012-2016).


\clearpage
\providecommand{\href}[2]{#2}\begingroup\raggedright\endgroup

\end{document}

%% file: IDM_table21.tex
\begin{tabular}{|c|c|c|c|c|c|}
%
%
\multicolumn{6}{c}{}\\
\multicolumn{6}{c}{a) Analysis of $e^+ e^- \to H^+ H^- \to W^+ W^- HH $}\\
\hline
Channel & Quantity & $\sqrt{s}$ [TeV] & BP1 & BP2 & BP3 \\
\hline
\multirow{2}{*}{$\lep\nu jjHH$} & \multirow{2}{*}{$m(j_1j_2)$ [GeV]} & 0.5 & 58.02 $\pm$ 0.10 & 47.11 $\pm$ 0.09 & 40.48 $\pm$ 0.30 \\
& & 1 & 59.17 $\pm$ 0.11 & 48.97 $\pm$ 0.09 & 42.32 $\pm$ 0.18 \\
\hline
\multirow{2}{*}{$jjjjHH$} & \multirow{2}{*}{$m(j_2j_3)$ [GeV]} & 0.5 & 59.68 $\pm$ 0.11 & 49.39 $\pm$ 0.12 & 39.94 $\pm$ 0.19 \\
& & 1 & 60.90 $\pm$ 0.10 & 50.45 $\pm$ 0.08 & 43.39 $\pm$ 0.22\\
\hline
\multirow{2}{*}{$jjjjHH$} & \multirow{2}{*}{$m(j_1j_4)$ [GeV]} & 0.5 & 59.71 $\pm$ 0.11 & 49.64 $\pm$ 0.11 & 40.25 $\pm$ 0.18 \\
& & 1 & 58.46 $\pm$ 0.15 & 48.48 $\pm$ 0.11 & 43.13 $\pm$ 0.21\\
\hline
\multicolumn{2}{|c|}{\multirow{2}{*}{Average $m_{H^{\pm}}-m_H$ [GeV]}} &
0.5 & 59.06 $\pm$ 0.06 & 48.44 $\pm$ 0.06 & 40.16 $\pm$ 0.12 \\
\multicolumn{2}{|c|}{} &
1  & 59.79 $\pm$ 0.07 & 49.50 $\pm$ 0.05 & 42.86 $\pm$ 0.12 \\
\hline
\hline
\multirow{2}{*}{$\lep\nu jjHH$} & \multirow{2}{*}{$\sum\limits_{i=1}^{2}E(j_i)$ [GeV]} & 0.5 & 123.16 $\pm$ 0.13 & 87.33 $\pm$ 0.11 & 58.68 $\pm$ 0.40 \\
& & 1 & 262.44 $\pm$ 0.22 & 190.95 $\pm$ 0.17 & 130.90 $\pm$ 0.31\\
\hline
\multirow{2}{*}{$jjjjHH$} & \multirow{2}{*}{$\sum\limits_{i=1}^{4}E(j_i)$ [GeV]} & 0.5 & 248.08 $\pm$ 0.17 & 175.85 $\pm$ 0.15 & 117.99 $\pm$ 0.31 \\
& & 1 & 525.46 $\pm$ 0.27 & 382.29 $\pm$ 0.22 & 261.85 $\pm$ 0.40\\
\hline
\multicolumn{2}{|c|}{Average $R=m_H/m_{H^{\pm}}$} &
0.5 & 50.49 $\pm$ 0.03 & 64.90 $\pm$ 0.02 & 76.42 $\pm$ 0.06 \\
\multicolumn{2}{|c|}{(in percent)} &
 1  & 47.47 $\pm$ 0.02 & 61.78 $\pm$ 0.02 & 73.82 $\pm$ 0.03 \\
\hline
%
%
\multicolumn{6}{c}{}\\
\multicolumn{6}{c}{b) Analysis of $e^+ e^- \to H A \to H H Z $}\\
\hline
\multirow{2}{*}{$\lep\lep HH$} & \multirow{2}{*}{$m(\lep_1\lep_2)$ [GeV]} & 0.5 & 55.37 $\pm$ 0.01 & 25.37 $\pm$ 0.01 & 5.86 $\pm$ 0.07 \\
& & 1 & 55.37 $\pm$ 0.01 & 25.37 $\pm$ 0.01 & 5.84 $\pm$ 0.03\\
\hline
\multirow{2}{*}{$jjHH$} & \multirow{2}{*}{$m(j_1j_2)$ [GeV]} & 0.5 & 49.21 $\pm$ 0.06 & 20.94 $\pm$ 0.03 & - \\
& & 1 & 49.58 $\pm$ 0.10 & 21.50 $\pm$ 0.05 & 4.64 $\pm$ 0.03\\
\hline
\multicolumn{2}{|c|}{\multirow{2}{*}{Average $m_A-m_H$ [GeV]}} &
0.5 & 55.20 $\pm$ 0.01 & 24.93 $\pm$ 0.01 & 5.86 $\pm$ 0.07 \\
\multicolumn{2}{|c|}{} &
 1  & 55.31 $\pm$ 0.01 & 25.22 $\pm$ 0.01 & 5.24 $\pm$ 0.02 \\
\hline
%
%
\multicolumn{6}{c}{}\\
\multicolumn{6}{c}{c) Reconstructed masses}\\
\hline
\multicolumn{2}{|c|}{\multirow{3}{*}{$m_{H^{\pm}}$ [GeV]}} & theo. & 123&140&176 \\
\multicolumn{2}{|c|}{}&0.5 & 119.29 $\pm$ 0.12 & 138.01 $\pm$ 0.17 & 170.31 $\pm$ 0.51 \\
\multicolumn{2}{|c|}{}& 1 & 113.82 $\pm$ 0.13 & 129.51 $\pm$ 0.13 & 163.71 $\pm$ 0.45 \\
\hline
\multicolumn{2}{|c|}{\multirow{3}{*}{$m_H$ [GeV]}} & theo. & 57.5&85.5&128 \\
\multicolumn{2}{|c|}{}&0.5 & 60.23 $\pm$ 0.06 & 89.57 $\pm$ 0.11 & 130.15 $\pm$ 0.39 \\
\multicolumn{2}{|c|}{}& 1 & 54.03 $\pm$ 0.06 & 80.01 $\pm$ 0.08 & 120.85 $\pm$ 0.33 \\
\hline
\multicolumn{2}{|c|}{\multirow{3}{*}{$m_A$ [GeV]}} & theo. & 113&111&134 \\
\multicolumn{2}{|c|}{}&0.5 & 115.43 $\pm$ 0.07 & 114.50 $\pm$ 0.12 & 136.01 $\pm$ 0.46 \\
\multicolumn{2}{|c|}{}& 1 & 109.34 $\pm$ 0.07 & 105.23 $\pm$ 0.09 & 126.09 $\pm$ 0.36 \\
\hline
\end{tabular}